%% file: main.tex
\documentclass[conference,compsoc]{IEEEtran}
\usepackage[justification=centering]{caption}

\usepackage{listings}
\usepackage{xurl}

\usepackage{graphicx}
\usepackage{cite}
\usepackage[table,xcdraw]{xcolor}
\usepackage{xspace}
\usepackage{multirow}
\usepackage{booktabs}
\usepackage{amsmath}
\usepackage{shortcuts} 
\usepackage{subcaption}
\usepackage{balance}

\usepackage{enumitem}
\setlist[itemize]{noitemsep, topsep=0pt}

\usepackage{amsfonts}
\usepackage{lipsum} 
\usepackage{placeins} 

\begin{document}
\title{\sys: A Benchmark for Realistic and Diverse Evaluation\\of Function Similarity in the Wild}
\author{
    \IEEEauthorblockN{Yiming Fan, Jun Yeon Won, Ding Zhu, Melih Sirlanci, Mahdi Khalili, Zhiqiang Lin, and Carter Yagemann}
    \IEEEauthorblockA{The Ohio State University\\\\
    fan.1192@osu.edu, won.126@osu.edu, zhu.3723@osu.edu, sirlanci.2@osu.edu,\\
    khalili.17@osu.edu, zlin@cse.ohio-state.edu, yagemann.1@osu.edu}
}


\maketitle

\pagestyle{plain}

\begin{abstract}
\input{paper/00_abs}
\end{abstract}

\section{Introduction}
\label{sec:introduction}
\input{paper/01_introduction}

\section{Background and Motivation}
\label{sec:background}
\input{paper/02_background}

\section{Dataset Design}
\label{sec:design}
\input{paper/03_dataset}

\section{Evaluation}
\label{sec:evaluation}
\input{paper/04_evaluation}

\section{Discussion}
\label{sec:discussion}
\input{paper/05_discussion}

\section{Conclusion}
\label{sec:conclusion}
\input{paper/06_conclusion}





\clearpage

\section*{Ethics Considerations}
\label{sec:ethics}
\input{paper/07_ethics}

\section*{Acknowledgments}
\label{sec:acknowledgments}
\input{paper/08_acknowledgments}
\section*{Availability}
\label{sec:availability}
\input{paper/09_availability}

\bibliographystyle{plain}
\bibliography{myReferences}

\end{document}

%% file: paper/00_abs.tex
Binary Function Similarity Detection (BFSD) is a core problem in software security, supporting tasks such as vulnerability analysis, malware classification, and patch provenance. In the past few decades, numerous models and tools have been developed for this application; however, due to the lack of a comprehensive universal benchmark in this field, researchers have struggled to compare different models effectively. Existing datasets are limited in scope, often focusing on a narrow set of transformations or type of binary, and fail to reflect the full diversity of real-world applications.


We introduce \sys, a benchmark comprising five realistic datasets collected from the wild, each highlighting a distinct aspect of the BFSD problem space. We evaluate \modelNumber representative models spanning multiple BFSD paradigms on \sys and observe performance degradations of up to $30\%$ on firmware and semantic datasets compared to standard settings, revealing substantial generalization gaps. Our results show that robustness to low- and mid-level binary variations does not generalize to high-level semantic differences, underscoring a critical blind spot in current BFSD evaluation practices.

%% file: paper/01_introduction.tex
\begin{figure}[!b]
\begin{center}
\includegraphics[width=\columnwidth]{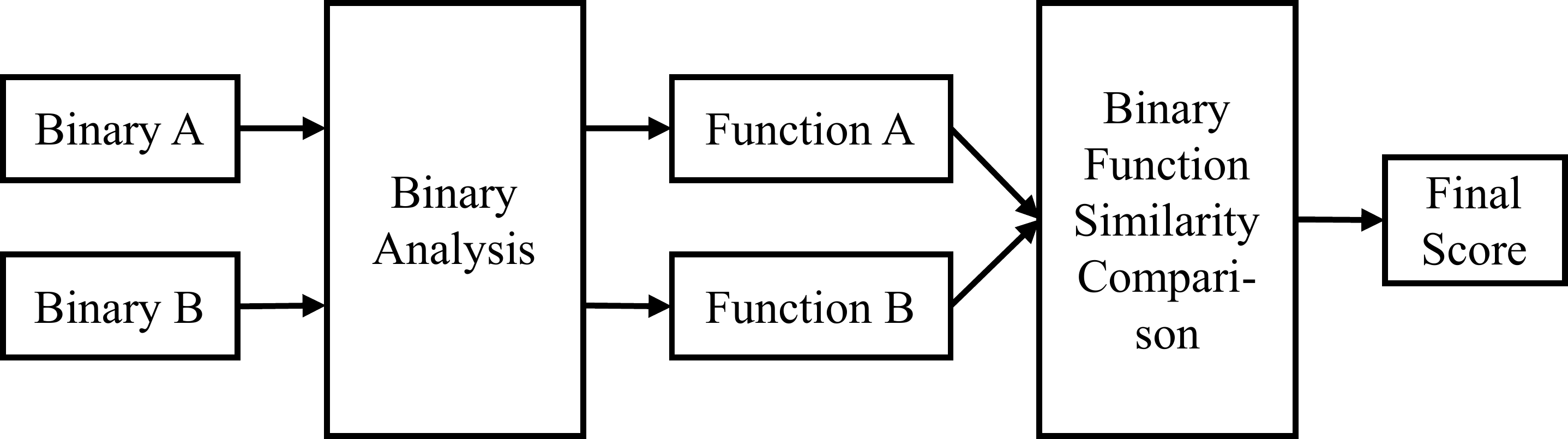}
\end{center}
\caption{\label{fig:problemDefinition}  Definition of the BFSD Problem }
\end{figure}

In binary function similarity detection (BFSD), the problem is framed as taking as input the binary form of two functions and generating an output that scores how similar they are\cite{usenix22}. Solutions to this problem can be applied in numerous security scenarios, such as bug searching\cite{SAFE,Asm2Vec}, where engineers compare unknown binaries to known ones to save effort. Other applications include malware clustering\cite{9110432}, detection\cite{deepbindiff, gemini} and lineage\cite{Genealogy}, patch analysis\cite{10.1145/3446371, Vuddy}, and vulnerability search\cite{ReDeBug}.

Despite extensive research, there remains unsolved challenges, one of which is the lack of a universal benchmark. Since this problem is applicable to many fields, including system security, software engineering, and machine learning, research is occurring in silos with no consensus on what makes a good evaluation dataset. This makes the findings harder to reproduce and compare. For example, some compiler optimization levels may produce little meaningful change in the resulting binary, artificially inflating metrics and limiting the dataset's effectiveness at differentiating models.

Existing datasets focus on covering different compiler versions and flags so that one source code produces many binaries, yielding multiple binary representations of the same function. However, researchers parameterize their frameworks differently, resulting in performance metrics that are not comparable or flawed. 

Another limitation of current datasets is that they sample from ``typical'' programs, introducing biases. For example, it is unknown whether comparing two \texttt{coreutils} functions is easier or harder than comparing firmware functions for cyberphysical systems. Most existing datasets cover the former, but not the latter\cite{usenix22}. The same is true for other code domains as well, such as malware. Although there exists malware datasets such as Malpedia \cite{malpedia}, these datasets are not built for the BFSD problem and lack clear function-level labels and controlled setups.


Finally, we lack a principled definition of how binary functions differ. Two semantically equivalent functions can vary substantially at the binary level. For example, compiling the same source code with different compilers produces a fundamentally different class of binary variation than compiling distinct but semantically equivalent source programs with the same compiler. Similarly, changing the optimization level typically introduces smaller differences than obfuscation techniques such as control-flow flattening or instruction substitution \cite{Obfuscator-LLVM}.

In this paper, we introduce \sys, the first truly diversified benchmark suite for binary function similarity detection (BFSD). We begin by defining a systematic three-level taxonomy of binary function differences, which provides a structured lens for understanding the sources of variation that challenge BFSD models:

\begin{figure*}[!t]
\begin{center}
\includegraphics[width=\textwidth]{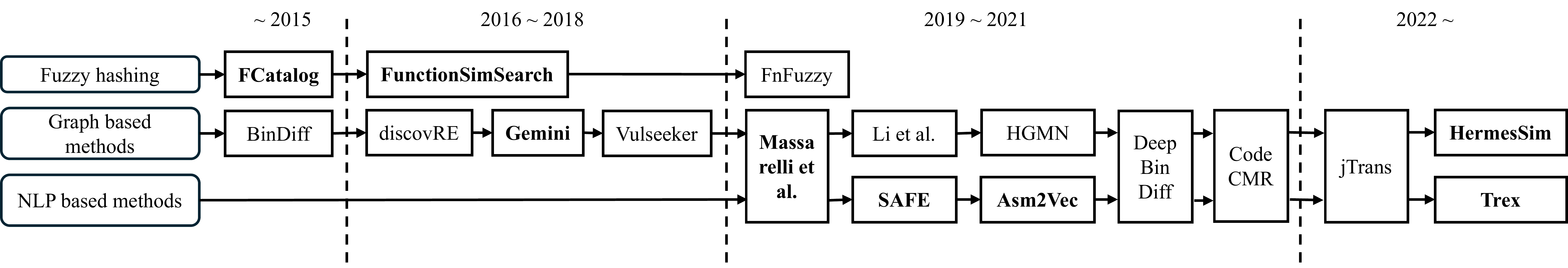}
\end{center}
\caption{\label{fig:systematization} Systematized Evolution Graph of the Three Main Trends of the BFSD Problem Solutions}
\end{figure*}

\begin{itemize}
\item \textit{Low-level differences}: variations in architecture, bitness, compilers, and optimization levels;
\item \textit{Mid-level differences}: obfuscation transformations such as control flow flattening and instruction substitution;
\item \textit{High-level differences}: independently developed implementations of the same functionality with substantially different source-level structure and control flow.
\end{itemize}

This taxonomy highlights a key limitation of prior benchmarks: most focus primarily on low-level recompilation settings, while mid- and high-level variations remain underexplored. Guided by this framework, we construct \sys to systematically cover all three categories. Beyond traditional recompiled open-source programs, \sys incorporates real-world firmware and malware binaries, state-of-the-art obfuscations, and semantically diverse programming-contest implementations.

By grounding dataset design in this taxonomy, \sys enables principled evaluation across distinct dimensions of binary variation and allows us to attribute performance degradation to specific levels of difference rather than treating robustness as a monolithic property.

Finally, with the datasets representing all the variations in the BFSD problem, we proceeded to evaluate \modelNumber\ models that represent the frontier landscape of research in this area, using \sys and addressing several interesting evaluation questions. The results reveal several notable trends in model performance across different datasets:

\begin{itemize} 
\item The AUC performances of the models are generally $10-20\%$ lower on the firmware dataset compared to the standard dataset. 
\item On the semantic dataset, models experience a more pronounced decline, with AUC performances typically $20-30\%$ lower than on the standard dataset. 
\item The obfuscation dataset also presents a challenge, leading to a moderate reduction in performance, with AUC scores about $5-10\%$ lower than those observed on the standard dataset.
\item HermesSim \cite{HermesSim}, the highest performing model on all the other datasets, failed to keep up with half of the models on the Semantic Dataset.
\end{itemize}

These findings highlight the importance of evaluating binary similarity models on diverse, realistic datasets, as performance can vary significantly depending on the characteristics of the binaries under consideration.

\paragraph{Contributions.} 
Our key contributions are as follows.
\begin{itemize} 
\item \textbf{Definition} of a three-level taxonomy of binary function differences, including low-level differences (architecture, compiler, optimization), mid-level differences (obfuscation techniques), and high-level differences (semantically equivalent but syntactically distinct functions).
\item \textbf{Development} of \sys, a diversified BFSD benchmark that systematically covers all three levels of variation within a unified evaluation framework, combining traditional recompiled code, real-world firmware and malware binaries, state-of-the-art obfuscation transformations, and semantically diverse implementations from programming contests.
\item \textbf{Evaluation} of leading BFSD models on \sys, revealing performance gaps of up to $30\%$ on semantic and firmware datasets and demonstrating that robustness claims derived from low-level settings do not generalize to high-level semantic variation.
\end{itemize}

%% file: paper/02_background.tex

In this section, we present a comprehensive picture of the problem and how researchers have been trying to solve it with a wide spectrum of methods. We also point out the drawbacks of the current datasets used to show our motivation.

\subsection{Problem Description}
\label{Problem Description}


Figure~\ref{fig:problemDefinition} illustrates the typical BFSD pipeline~\cite{gemini,SAFE}. Given two binaries $(A, B)$, a binary analysis stage extracts candidate functions; the BFSD model then evaluates pairs $(f_A, f_B)$ and outputs a similarity score. This score can be used either for threshold-based match/no-match decisions or for similarity ranking in retrieval settings, where functions are ranked by similarity to a query~\cite{Asm2Vec,deepbindiff}. While this depiction captures the conceptual workflow, real-world systems vary substantially in their analysis front-ends, feature extraction strategies, and matching objectives.

\paragraph{Binary Analysis.} Researchers have been using different binary analysis tools and different forms of binary functions as input to the comparison model. The raw bytes of the binaries is the simplest binary form, which some methods take directly as input\cite{FCatalog}. To achieve better comparison results, researchers usually go further and use binary analysis tools such as IDA Pro \cite{IDAPro} or Ghidra \cite{ghidra} to extract more features; for instance, some approaches use assembly \cite{Asm2Vec, Trex} or normalized assembly\cite{SAFE}, while others use intermediate representation \cite{10.1145/3052973.3052995, VulSeeker, Yu_Cao_Tang_Nie_Huang_Wu_2020} or Control-Flow Graphs (CFGs) \cite{gemini, 10.1145/2976749.2978370, discovRE, BinSequence}. Moreover, others have even tried using dynamic analysis \cite{Blanket_execution, 10.5555/3155562.3155608, BinSim, 10.1145/3634737.3644996} and symbolic execution \cite{BinGo}. There are even methods that combine dynamic execution with static emulation of target functions \cite{binmatch}.

\paragraph{Binary Function Similarity Comparison.} Some solutions need to compare function pairs one by one directly to calculate their similarity score. Because they usually take in two raw or processed binary functions and output a result, this type of work is very suitable for machine learning. Researchers have applied numerous machine learning techniques, such as Convolutional
Neural Networks (CNN) \cite{10.1145/3238147.3238199} and Graph Matching Networks (GMN) \cite{10376642}, among other neural networks\cite{Zeek}.

\subsection{Methods Systematization}
\label{Methods Systematization}

\input{tables/datasetcomparison}

Systematizing the BFSD literature is inherently difficult as the work spans multiple areas. A full survey is beyond our scope; we offer only a brief overview and a simple three-trend taxonomy to frame our study.

\paragraph{Fuzzy hashing}, also referred to as similarity hashing, aims to generate hash values that preserve similarity between inputs, such that similar binaries yield comparable hash outputs. Early schemes such as ssdeep~\cite{Kornblum2006} and sdhash~\cite{Roussev2010} have been widely used in malware analysis and approximate binary matching. However, prior studies have shown that traditional fuzzy hashing applied directly to raw binary bytes can be overly sensitive to minor modifications, limiting its robustness in the presence of compiler optimizations or small code changes~\cite{10.1145/3176258.3176306}. To mitigate this issue, subsequent work introduced feature-based or structure-aware hashing techniques that incorporate program-level features prior to hashing, improving resilience and retrieval performance~\cite{functionsimsearch}.

\paragraph{Graph based methods} typically take advantage of the graphic structures extracted from the binaries, e.g., intra-procedural Control Flow Graphs (CFG). A well-known graph-based differencing tool is BinDiff \cite{bindiff}, which uses CFG normalization and approximate graph isomorphism to match functions across binaries. Another classic method that integrates CFG analysis is Kam1n0 \cite{Kam1n0}, a scalable assembly code clone search engine designed to assist reverse engineers in identifying similar or cloned binary functions within large codebases. Machine learning techniques like Graph Neural Network (GNN) were also employed. Many variations of GNN ensued from the success of these methods. For instance, Li et al. \cite{lielal} proposed a solution based on Graph Matching Network (GMN), which takes as input two graphs and calculates their similarity. This work treats BFSD simply as one specific application of GMN. Other improvements in these types of methods include creating new forms of graphs that can better represent the semantic contents of the binary functions; for example, HermesSim \cite{HermesSim} came up with a new representation of binary functions named Semantics-Oriented Graph (SOG) by tokenizing intra-instruction structures and eliminating tokens irrelevant to the semantic contents.

\paragraph{NLP based methods}are usually adapted from existing Natural Language Processing (NLP) techniques and solve the BFSD problem by simply treating assembly code as text. Within this main trend, there are three smaller streams based on three specific NLP techniques. The first one is word2vec \cite{Mikolov2013EfficientEO, NIPS2013_9aa42b31}, which can calculate continuous vector representations of words from large datasets. This technique inspired Asm2Vec \cite{Asm2Vec}, which converts the assembly form of binary functions into code embeddings. It works on the instruction level to avoid the Out-of-Vocabulary (OOV) problem that bugs many similar approaches. The second stream of solutions are inspired by seq2seq \cite{seq2seq}, which utilizes multiple Long Short-Term Memory (LSTM) layers to map sequences to vectors, and back to the target sequences. This technique makes it possible to map binary functions from different architectures to the same embedding space, enabling solutions with cross-architecture capabilities. Following seq2seq, solutions such as SAFE \cite{SAFE} mushroomed. The last technique is transformer \cite{transformer}, which led to the development of BERT \cite{BERT}, a very popular pre-training model in the NLP area. Many solutions are based on BERT, such as Yu et al. \cite{Yu_Cao_Tang_Nie_Huang_Wu_2020}. Other models like Trex \cite{Trex} and jTrans \cite{jTrans} also consist of Transformer-based language models.

\paragraph{LLM-Based Methods.}
Recent work has begun incorporating pretrained language models and large language models (LLMs) into binary similarity analysis. CLAP~\cite{CLAP} uses natural-language supervision to learn transferable binary code representations, while Co\textsuperscript{2}FuLL~\cite{Co2FuLL} combines conventional similarity analysis with LLM-based candidate verification and explanation. These approaches extend BFSD beyond task-specific representation learning by leveraging pretrained models and broader semantic information. However, their evaluation also introduces new considerations, including dependence on model versions, prompting strategies, and potential overlap between benchmark data and model pretraining corpora. \sys does not currently evaluate LLM-based BFSD methods, and we leave systematic evaluation under contamination-controlled settings to future work.

\paragraph{Other notable methods} include DeepBinDiff and BinPro. DeepBinDiff learns semantic/structural embeddings of basic blocks via Text-Associated DeepWalk and aligns functions with a k-hop greedy matcher~\cite{deepbindiff}; whereas BinPro targets source-to-binary provenance by extracting features from binaries and source (e.g., constants, call graphs) and using ML to predict similarity, remaining robust to compiler/optimization changes~\cite{binpro}.

As sketched in Figure~\ref{fig:systematization}, the field progressed from early fuzzy hashing to graph-based methods and, more recently, to NLP-based approaches often combined with graphs. The figure is illustrative (not exhaustive), and a full systematization is outside our scope.

\subsection{Motivation}
\label{Motivation}

As we perused through the research papers that deal with BFSD, we can see that the datasets they used are too heterogeneous for readers to actually compare multiple solutions horizontally. To demonstrate this, we list some of the most important and state-of-the-art works here and compare the datasets they used in Table~\ref{tab:comparison}, where the datasets are classified into five categories, namely, standard, firmware, malware, obfuscated, and semantic dataset. And we use letters to represent the variables they used to create the differences inside the binary function pairs, as described under the table. Empty circle represents an empty set, while half-circles are strict subsets of the full circle in the same column.

We can see that most previous research on BFSD used only the standard dataset, e.g. \texttt{OpenSSL} and \texttt{Binutils}. Only two studies tested their models on a dataset created by using different obfuscation techniques \cite{Asm2Vec,Trex}. Very few studies included firmware binaries and malware binaries in their dataset and none of the studies tried using a semantic dataset compiled from different source codes that implement the same function.

Even when we only consider the standard binary dataset, previous evaluations lack diversity in the variables used to create the difference in binary function pairs. All studies include the optimization level as a variable; however, few considered bitness and other variables such as compiler version and architecture. One might think that as time elapses, researchers will tend to improve and enrich their dataset, but being published in 2022, jTrans \cite{jTrans} only considered the optimization level in their dataset design, which made it significantly more difficult for the following studies to compare performance with their model\cite{HermesSim}.


%% file: tables/datasetcomparison.tex
\begin{table*}[!t]
\footnotesize
\centering
\caption{A Comprehensive Comparison of the Datasets Used in Binary Function Similarity Studies.}
\label{tab:comparison}
\scalebox{1}{
\begin{tblr}{
  column{3} = {c},
  column{4} = {c},
  column{5} = {c},
  column{6} = {c},
  column{7} = {c},
  column{9} = {c},
  column{10} = {c},
  column{12} = {c},
  column{13} = {c},
  column{14} = {c},
  column{16} = {c},
  column{18} = {c},
  cell{1}{1} = {r=2}{},
  cell{1}{3} = {c=5}{},
  cell{1}{9} = {c=2}{},
  cell{1}{12} = {c=3}{},
  cell{3}{8} = {c},
  cell{3}{11} = {c},
  cell{3}{15} = {c},
  cell{3}{17} = {c},
  cell{4}{8} = {c},
  cell{4}{11} = {c},
  cell{4}{15} = {c},
  cell{4}{17} = {c},
  cell{5}{8} = {c},
  cell{5}{11} = {c},
  cell{5}{15} = {c},
  cell{5}{17} = {c},
  cell{6}{8} = {c},
  cell{6}{11} = {c},
  cell{6}{15} = {c},
  cell{6}{17} = {c},
  cell{7}{8} = {c},
  cell{7}{11} = {c},
  cell{7}{15} = {c},
  cell{7}{17} = {c},
  cell{8}{8} = {c},
  cell{8}{11} = {c},
  cell{8}{15} = {c},
  cell{8}{17} = {c},
  cell{9}{8} = {c},
  cell{9}{11} = {c},
  cell{9}{15} = {c},
  cell{9}{17} = {c},
  cell{10}{8} = {c},
  cell{10}{11} = {c},
  cell{10}{15} = {c},
  cell{10}{17} = {c},
  cell{11}{8} = {c},
  cell{11}{11} = {c},
  cell{11}{15} = {c},
  cell{11}{17} = {c},
  cell{12}{8} = {c},
  cell{12}{11} = {c},
  cell{12}{15} = {c},
  cell{12}{17} = {c},
  hline{1,3,13} = {-}{},
  hline{2} = {3-7,9-10,12-14,16,18}{},
}
Work Name                           &  & Standard   &            &            &            &           &  & Firmware   &            &  & Malware    &            &            &  & Obfuscated &  & Semantic   \\
                                    &  & A          & B          & C          & V          & O         &  & V          & O          &  & C          & V          & O          &  & Ob         &  & S          \\
Zeek\cite{Zeek}                     &  & \halfcirc  & \emptycirc & \halfcirc  & \halfcirc  & \halfcirc &  & \emptycirc & \emptycirc &  & \emptycirc & \emptycirc & \emptycirc &  & \emptycirc &  & \emptycirc \\
Massarelli et al. \cite{Massarelli} &  & \halfcirc  & \emptycirc & \halfcirc  & \halfcirc  & \halfcirc &  & \emptycirc & \emptycirc &  & \emptycirc & \emptycirc & \emptycirc &  & \emptycirc &  & \emptycirc \\
Li et al. \cite{lielal}             &  & \emptycirc & \emptycirc & \halfcirc  & \emptycirc & \halfcirc &  & \emptycirc & \emptycirc &  & \emptycirc & \emptycirc & \emptycirc &  & \emptycirc &  & \emptycirc \\
SAFE\cite{SAFE}                     &  & \halfcirc  & \emptycirc & \halfcirc  & \halfcirc  & \halfcirc &  & \emptycirc & \emptycirc &  & \emptycirc & \emptycirc & \emptycirc &  & \emptycirc &  & \emptycirc \\
Asm2Vec\cite{Asm2Vec}               &  & \emptycirc & \emptycirc & \emptycirc & \emptycirc & \halfcirc &  & \emptycirc & \emptycirc &  & \emptycirc & \emptycirc & \emptycirc &  & \halfcirc  &  & \emptycirc \\
Marcelli et al.\cite{usenix22}      &  & \halfcirc  & \halfcirc  & \halfcirc  & \halfcirc  & \halfcirc &  & \emptycirc & \emptycirc &  & \emptycirc & \emptycirc & \emptycirc &  & \emptycirc &  & \emptycirc \\
jTrans\cite{jTrans}                 &  & \emptycirc & \emptycirc & \emptycirc & \emptycirc & \halfcirc &  & \emptycirc & \emptycirc &  & \emptycirc & \emptycirc & \emptycirc &  & \emptycirc &  & \emptycirc \\
Trex\cite{Trex}                     &  & \halfcirc  & \emptycirc & \emptycirc & \emptycirc & \halfcirc &  & \emptycirc & \emptycirc &  & \emptycirc & \emptycirc & \emptycirc &  & \halfcirc  &  & \emptycirc \\
HermesSim\cite{HermesSim}           &  & \halfcirc  & \halfcirc  & \halfcirc  & \halfcirc  & \halfcirc &  & \emptycirc & \emptycirc &  & \emptycirc & \emptycirc & \emptycirc &  & \emptycirc &  & \emptycirc \\
\sys                                &  & \fullcirc  & \fullcirc  & \fullcirc  & \fullcirc  & \fullcirc &  & \fullcirc  & \fullcirc  &  & \fullcirc  & \fullcirc  & \fullcirc  &  & \fullcirc  &  & \fullcirc  
\end{tblr}
}
\captionsetup{justification=centering, margin=0.5cm}
\caption*{\small
A: Architecture; B: Bitness; C: Compiler; V: Compiler Version;
O: Optimization Level; Ob: Obfuscation; S: Semantics.
}
\end{table*}
\vspace{-0.5\baselineskip}

%% file: paper/03_dataset.tex
\input{tables/tasks}

Motivated by the shortcomings of existing datasets, we construct a more diverse dataset covering different binary types and a broader range of variables that introduce differences within binary function pairs, as summarized in Table~\ref{tab:tasks}. In the table, rows represent tasks and columns represent the corresponding variables, with checkmarks indicating which variables are varied in each task. To prevent information leakage, we perform train/validation/test splitting at the source-entity level rather than at the binary or function level. For the Semantic Dataset, splitting is performed at the submission level, whereas all other datasets are split at the project level, ensuring that binaries and functions from the same source entity remain within the same split.


\paragraph{The Standard Dataset} comprises popular open-source projects that reflect the complexity of real-world software. To ensure the comprehensiveness of the standard dataset, we gathered all open source projects used in previous research and added more, such as \texttt{ffmpeg}, \texttt{gettext}, \texttt{curl} and \texttt{ImageMagick}. In addition to that, we also included command line utility libraries, for instance, \texttt{binutils}, \texttt{coreutils}, \texttt{diffutils}, \texttt{findutils}, etc.

For the standard dataset, we used architecture, bitness, compiler, compiler version, and optimization level as variables to create the difference within binary function pairs. Specifically, we utilized 3 different architectures, namely ARM, MIPS, and x86-64, with 32-bit and 64-bit variations. For the compilation process, we used two compilers, each with 4 different versions (\texttt{clang 3.5/5/7/9} and \texttt{GCC 4.8/5/7/9}), and the compilation was implemented at 5 different optimization levels (\texttt{O0}, \texttt{O1}, \texttt{O2}, \texttt{O3}, \texttt{Os}). 

As practiced in several seminal papers such as \cite{discovRE}, we followed the rule in our function pair generation process to filter out functions with fewer than five basic blocks and function pairs with the same content hash to ensure there is enough complexity in the binary function pairs. The size of each dataset in terms of library/binary/function numbers can be found in Table~\ref{tab:dataset_size}.

\input{tables/datasetSize}


\paragraph{The Firmware Dataset} is generated with several hardware vendors and their corresponding Integrated Development Environments (IDEs) based on source code availability and command-line compatibility, namely Keil MDK-ARM~\cite{KeilMDK} for Nuvoton, Simplicity Studio~\cite{SimplicityStudio} for Silicon Labs, and nRF5 SDK~\cite{nrf5_sdk} for Nordic Semiconductor. The resulting firmwares implement a wide range of functionality, including wireless communication stacks (BLE, Mesh, Zigbee, Thread, Wi-SUN, ANT, NFC, and proprietary 2.4 GHz protocols), peripheral and sensor drivers (UART, SPI, I²C, ADC/DAC, timers, motor control), and system services (bootloaders, in-system programming, RTOS integration), as well as demonstration applications for signal processing, energy management, and device profiling. These reference implementations capture the connectivity, sensing, and control capabilities of modern microcontrollers, providing a diverse basis for evaluating binary function similarity models.

Due to the nature of the firmware binaries, we were unable to freely utilize variables such as architecture, bitness, or different compilers to create the difference within binary function pairs. As a result, we used 15 versions of \texttt{ARM} compiler and compiled the firmware source codes at 6 different optimization levels (\texttt{O0}, \texttt{O1}, \texttt{O2}, \texttt{O3}, \texttt{Os}, \texttt{Ofast}).  


\paragraph{The Malware Dataset} source code come from two public repositories, the MalwareSourceCode repository~\cite{vxunderground_malware} and the IoT Malware repository~\cite{iot-malware-2017, ding2020deeppower}. These repositories contain a wide range of Linux malware, including kernel and userland rootkits (e.g., Adore-ng, Azazel, Jynx), IoT botnets (e.g., Mirai variants, Kaiten, BASHLITE, LizardStresser), and smaller families such as IRC bots, worms, and ransomware. Some families appear with multiple forks or variants (e.g., Mirai, the B1NARY series, Jynx), while others are represented by a single project, giving our dataset both breadth and depth across different malware types. 

To create differences within the binary function pairs, we compiled the source codes using multiple compilers, compiler versions, and optimization levels. In particular, we used five versions of \texttt{Clang} (11, 12, 13, 14, and 15) and four versions of \texttt{GCC} (9, 10, 11, and 12), each at five optimization levels (\texttt{O0}, \texttt{O1}, \texttt{O2}, \texttt{O3}, and \texttt{Os}).  

Because it is relatively difficult to find compilable malware source code in the C language, the malware dataset is not the largest dataset we built. Nevertheless, malware authors often employ obfuscation techniques to protect and disguise their payloads. Consequently, we expect the performance of current solutions on malware binaries to be particularly interesting compared with the standard dataset.


\input{tables/generalEvaluation}

\paragraph{The Obfuscation Dataset.} Code obfuscation is a well-known obstacle for BFSD solutions. While some prior work has considered it when constructing binary function pairs~\cite{Asm2Vec, Trex}, many datasets omit it, leaving its impact unclear. To address this gap, we built an obfuscation dataset from open-source projects and command-line utility libraries from the standard dataset, with code obfuscation being the only variable used to create differences within function pairs. 

We used three obfuscators: Obfuscator-LLVM~\cite{Obfuscator-LLVM}, Hikari~\cite{hikari}, and Tigress~\cite{tigress}. For Obfuscator-LLVM, we employed four configurations, namely \texttt{Instruction Substitution}, \texttt{Bogus Control Flow}, \texttt{Control Flow Flattening}, and the combination of all three. For Hikari, since the project is deprecated and no longer maintained, we incorporated the obfuscation dataset used in Trex~\cite{Trex}, which includes five techniques: \texttt{Bogus Control Flow}, \texttt{Control Flow Flattening}, \texttt{Register-Based Indirect Branching}, \texttt{Basic Block Splitting}, and \texttt{Instruction Substitution}. Tigress is a source-level obfuscation framework that supports a wide range of transformations, including virtualization-based obfuscation. In this work, we apply five transformations: \texttt{Add Opaque Predicates}, \texttt{Add Stack Variables}, \texttt{Encode Arithmetic and Data}, \texttt{Initialize Opaque Variables}, and \texttt{Reorder Functions and Blocks}. These transformations are chosen because they are distinctive to Tigress, widely studied, and preserve function boundaries, ensuring the ground truth mapping between functions remains valid.    


\paragraph{The Semantic Dataset.} The first four datasets evaluate variations introduced by changing compilation variables while preserving a shared source code base. In contrast, the semantic dataset targets high-level differences in our taxonomy, where independent implementations realize the same functionality but differ substantially in source-level structure, control flow organization, and data abstractions.

To construct this dataset at scale, we collect source code from a programming contest platform~\cite{atcoder}, where multiple participants independently implement solutions to the same problem specification. Submissions under the same problem are designed to satisfy identical functional requirements while differing syntactically and structurally. We leverage this property to approximate semantic equivalence across independently written programs.

Specifically, we gather all C-language submissions from 16 weekly contests spanning six difficulty levels to ensure diversity in complexity. All selected submissions passed the official judge of the contest platform, ensuring functional equivalence under standardized test inputs. To reduce labeling noise and isolate semantically comparable functions, we filter out submissions that rely on auxiliary helper functions and focus on the main implementation function in each program.

Positive function pairs are formed by pairing two submissions from the same contest problem, while negative pairs are constructed from submissions corresponding to different problems. This dataset therefore provides a large-scale evaluation of BFSD robustness under high-level semantic differences that are not induced by compiler settings or obfuscation, but by independent program design.

%% file: tables/tasks.tex
\begin{table}[t]
\scriptsize
\centering
\caption{Variables Used in Different Tasks.}
\label{tab:tasks}

\begin{tblr}{
  column{even} = {c},
  column{3} = {c},
  column{5} = {c},
  column{7} = {c},
  hline{1-2,12} = {-}{},
}
Task \textbackslash Variables & A & B & C & V & O & Ob & S \\

XA          & \cmark       & \xmark  & \xmark   & \xmark            & \xmark       & \xmark      & \xmark   \\
XB          & \xmark       & \cmark  & \xmark   & \xmark            & \xmark       & \xmark      & \xmark   \\
XA+XB       & \cmark       & \cmark  & \xmark   & \xmark            & \xmark       & \xmark      & \xmark   \\
XC          & \xmark       & \xmark  & \cmark   & \cmark            & \cmark       & \xmark      & \xmark   \\
XC+XB       & \xmark       & \cmark  & \cmark   & \cmark            & \cmark       & \xmark      & \xmark   \\
XV          & \xmark       & \xmark  & \xmark   & \cmark            & \xmark       & \xmark      & \xmark   \\
XO          & \xmark       & \xmark  & \xmark   & \xmark            & \cmark       & \xmark      & \xmark   \\
XM          & \cmark       & \cmark  & \cmark   & \cmark            & \cmark       & \xmark      & \xmark   \\
XOb         & \xmark       & \xmark  & \xmark   & \xmark            & \xmark       & \cmark      & \xmark   \\
XS          & \xmark       & \xmark  & \xmark   & \xmark            & \xmark       & \xmark      & \cmark   
\end{tblr}

\end{table}
\vspace{-0.5\baselineskip}

%% file: tables/datasetSize.tex
\begin{table}
\centering
\caption{Overall Dataset Size and Splits.}
\label{tab:dataset_size}
\scalebox{0.6}{
\begin{tblr}{
  row{1} = {c},
  cell{2}{2} = {r},
  cell{2}{3} = {r},
  cell{2}{4} = {r},
  cell{2}{5} = {r},
  cell{2}{6} = {r},
  cell{2}{7} = {r},
  cell{2}{8} = {r},
  cell{2}{9} = {r},
  cell{2}{10} = {r},
  cell{2}{11} = {r},
  cell{2}{12} = {r},
  cell{3}{2} = {r},
  cell{3}{3} = {r},
  cell{3}{4} = {r},
  cell{3}{5} = {r},
  cell{3}{6} = {r},
  cell{3}{7} = {r},
  cell{3}{8} = {r},
  cell{3}{9} = {r},
  cell{3}{10} = {r},
  cell{3}{11} = {r},
  cell{3}{12} = {r},
  cell{4}{2} = {r},
  cell{4}{3} = {r},
  cell{4}{4} = {r},
  cell{4}{5} = {r},
  cell{4}{6} = {r},
  cell{4}{7} = {r},
  cell{4}{8} = {r},
  cell{4}{9} = {r},
  cell{4}{10} = {r},
  cell{4}{11} = {r},
  cell{4}{12} = {r},
  cell{5}{2} = {r},
  cell{5}{3} = {r},
  cell{5}{4} = {r},
  cell{5}{5} = {r},
  cell{5}{6} = {r},
  cell{5}{7} = {r},
  cell{5}{8} = {r},
  cell{5}{9} = {r},
  cell{5}{10} = {r},
  cell{5}{11} = {r},
  cell{5}{12} = {r},
  cell{6}{2} = {r},
  cell{6}{3} = {r},
  cell{6}{4} = {r},
  cell{6}{5} = {r},
  cell{6}{6} = {r},
  cell{6}{7} = {r},
  cell{6}{8} = {r},
  cell{6}{9} = {r},
  cell{6}{10} = {r},
  cell{6}{11} = {r},
  cell{6}{12} = {r},
  cell{7}{2} = {r},
  cell{7}{3} = {r},
  cell{7}{4} = {r},
  cell{7}{5} = {r},
  cell{7}{6} = {r},
  cell{7}{7} = {r},
  cell{7}{8} = {r},
  cell{7}{9} = {r},
  cell{7}{10} = {r},
  cell{7}{11} = {r},
  cell{7}{12} = {r},
  hline{1-2,8} = {-}{},
}
Dataset      &  & Standard  &  & Firmware &  & Malware &  & Obfuscated &  & Semantic &  \\
Libraries    &  & 181       &  & 159      &  & 61      &  & 125        &  & 75       &  \\
Binaries     &  & 16,889    &  & 6,086    &  & 2,905   &  & 4,775      &  & 5,415    &  \\
Functions    &  & 1,601,164 &  & 19,794   &  & 6,225   &  & 139,045    &  & 4,165    &  \\
~-Training   &  & 1,249,747 &  & 11,302   &  & 3,669   &  & 92,972     &  & 2,322    &  \\
~-Validation &  & 41,893    &  & 795      &  & 1,118   &  & 16,823     &  & 592      &  \\
~-Testing    &  & 309,524   &  & 7,697    &  & 1,438   &  & 29,250     &  & 1,251    &  
\end{tblr}
}
\end{table}

%% file: tables/generalEvaluation.tex
\begin{table*}[!t]
\centering
\caption{An Overview of the AUC Comparison on All Datasets.}
\label{tab:AUC_overview}
\scalebox{0.8}{
\begin{tblr}{
  row{2} = {c},
  cell{1}{1} = {r=2}{},
  cell{1}{2} = {c=6}{c},
  cell{1}{8} = {c},
  cell{1}{9} = {c=2}{c},
  cell{1}{11} = {c},
  cell{1}{12} = {c=3}{c},
  cell{1}{15} = {c},
  cell{1}{16} = {c},
  cell{1}{17} = {c},
  cell{1}{18} = {c},
  cell{1}{19} = {c},
  cell{3}{2} = {c},
  cell{3}{3} = {c},
  cell{3}{4} = {c},
  cell{3}{5} = {c},
  cell{3}{6} = {c},
  cell{3}{7} = {c},
  cell{3}{8} = {c},
  cell{3}{9} = {c},
  cell{3}{10} = {c},
  cell{3}{11} = {c},
  cell{3}{12} = {c},
  cell{3}{13} = {c},
  cell{3}{14} = {c},
  cell{3}{15} = {c},
  cell{3}{16} = {c},
  cell{3}{17} = {c},
  cell{3}{18} = {c},
  cell{3}{19} = {c},
  cell{4}{2} = {c},
  cell{4}{3} = {c},
  cell{4}{4} = {c},
  cell{4}{5} = {c},
  cell{4}{6} = {c},
  cell{4}{7} = {c},
  cell{4}{8} = {c},
  cell{4}{9} = {c},
  cell{4}{10} = {c},
  cell{4}{11} = {c},
  cell{4}{12} = {c},
  cell{4}{13} = {c},
  cell{4}{14} = {c},
  cell{4}{15} = {c},
  cell{4}{16} = {c},
  cell{4}{17} = {c},
  cell{4}{18} = {c},
  cell{4}{19} = {c},
  cell{5}{2} = {c},
  cell{5}{3} = {c},
  cell{5}{4} = {c},
  cell{5}{5} = {c},
  cell{5}{6} = {c},
  cell{5}{7} = {c},
  cell{5}{8} = {c},
  cell{5}{9} = {c},
  cell{5}{10} = {c},
  cell{5}{11} = {c},
  cell{5}{12} = {c},
  cell{5}{13} = {c},
  cell{5}{14} = {c},
  cell{5}{15} = {c},
  cell{5}{16} = {c},
  cell{5}{17} = {c},
  cell{5}{18} = {c},
  cell{5}{19} = {c},
  cell{6}{2} = {c},
  cell{6}{3} = {c},
  cell{6}{4} = {c},
  cell{6}{5} = {c},
  cell{6}{6} = {c},
  cell{6}{7} = {c},
  cell{6}{8} = {c},
  cell{6}{9} = {c},
  cell{6}{10} = {c},
  cell{6}{11} = {c},
  cell{6}{12} = {c},
  cell{6}{13} = {c},
  cell{6}{14} = {c},
  cell{6}{15} = {c},
  cell{6}{16} = {c},
  cell{6}{17} = {c},
  cell{6}{18} = {c},
  cell{6}{19} = {c},
  cell{7}{2} = {c},
  cell{7}{3} = {c},
  cell{7}{4} = {c},
  cell{7}{5} = {c},
  cell{7}{6} = {c},
  cell{7}{7} = {c},
  cell{7}{8} = {c},
  cell{7}{9} = {c},
  cell{7}{10} = {c},
  cell{7}{11} = {c},
  cell{7}{12} = {c},
  cell{7}{13} = {c},
  cell{7}{14} = {c},
  cell{7}{15} = {c},
  cell{7}{16} = {c},
  cell{7}{17} = {c},
  cell{7}{18} = {c},
  cell{7}{19} = {c},
  cell{8}{2} = {c},
  cell{8}{3} = {c},
  cell{8}{4} = {c},
  cell{8}{5} = {c},
  cell{8}{6} = {c},
  cell{8}{7} = {c},
  cell{8}{8} = {c},
  cell{8}{9} = {c},
  cell{8}{10} = {c},
  cell{8}{11} = {c},
  cell{8}{12} = {c},
  cell{8}{13} = {c},
  cell{8}{14} = {c},
  cell{8}{15} = {c},
  cell{8}{16} = {c},
  cell{8}{17} = {c},
  cell{8}{18} = {c},
  cell{8}{19} = {c},
  cell{9}{2} = {c},
  cell{9}{3} = {c},
  cell{9}{4} = {c},
  cell{9}{5} = {c},
  cell{9}{6} = {c},
  cell{9}{7} = {c},
  cell{9}{8} = {c},
  cell{9}{9} = {c},
  cell{9}{10} = {c},
  cell{9}{11} = {c},
  cell{9}{12} = {c},
  cell{9}{13} = {c},
  cell{9}{14} = {c},
  cell{9}{15} = {c},
  cell{9}{16} = {c},
  cell{9}{17} = {c},
  cell{9}{18} = {c},
  cell{9}{19} = {c},
  cell{10}{2} = {c},
  cell{10}{3} = {c},
  cell{10}{4} = {c},
  cell{10}{5} = {c},
  cell{10}{6} = {c},
  cell{10}{7} = {c},
  cell{10}{8} = {c},
  cell{10}{9} = {c},
  cell{10}{10} = {c},
  cell{10}{11} = {c},
  cell{10}{12} = {c},
  cell{10}{13} = {c},
  cell{10}{14} = {c},
  cell{10}{15} = {c},
  cell{10}{16} = {c},
  cell{10}{17} = {c},
  cell{10}{18} = {c},
  cell{10}{19} = {c},
  cell{11}{2} = {c},
  cell{11}{3} = {c},
  cell{11}{4} = {c},
  cell{11}{5} = {c},
  cell{11}{6} = {c},
  cell{11}{7} = {c},
  cell{11}{8} = {c},
  cell{11}{9} = {c},
  cell{11}{10} = {c},
  cell{11}{11} = {c},
  cell{11}{12} = {c},
  cell{11}{13} = {c},
  cell{11}{14} = {c},
  cell{11}{15} = {c},
  cell{11}{16} = {c},
  cell{11}{17} = {c},
  cell{11}{18} = {c},
  cell{11}{19} = {c},
  hline{1,3,12} = {-}{},
  hline{2} = {2-7,9-10,12-14,16,18}{},
}
Model Name                                 & Standard      &      &       &               &               &      &  & Firmware      &      &  & Malware &               &      &  & Obfuscated    &  & Semantic &  \\
                                           & XA            & XC   & XC+XB & XV            & XO            & XM   &  & XV            & XO   &  & XC      & XV            & XO   &  & XOb           &  & XS       &  \\
FCatalog \cite{FCatalog}                   & 0.44          & 0.81 & 0.62  & \textbf{0.99} & 0.94          & 0.66 &  & \textbf{0.99} & 0.90 &  & 0.83    & \textbf{0.99} & 0.89 &  & 0.78          &  & 0.70     &  \\
FunctionSimSearch \cite{functionsimsearch} & 0.72          & 0.66 & 0.65  & 0.81          & 0.79          & 0.72 &  & 0.74          & 0.61 &  & 0.80    & \textbf{0.97} & 0.90 &  & 0.64          &  & 0.68     &  \\
Asm2Vec \cite{Asm2Vec}                     & 0.44          & 0.71 & 0.62  & 0.81          & 0.77          & 0.70 &  & 0.62          & 0.57 &  & 0.92    & \textbf{0.99} & 0.91 &  & 0.67          &  & 0.53     &  \\
Massarelli et al. \cite{Massarelli}        & 0.87          & 0.80 & 0.78  & \textbf{0.95} & 0.90          & 0.88 &  & 0.91          & 0.80 &  & 0.90    & 0.91          & 0.93 &  & 0.92          &  & 0.73     &  \\
Gemini \cite{gemini}                       & 0.96          & 0.80 & 0.79  & 0.98          & 0.91          & 0.90 &  & 0.96          & 0.85 &  & 0.89    & \textbf{0.99} & 0.95 &  & 0.84          &  & 0.83     &  \\
SAFE \cite{SAFE}                           & 0.79          & 0.83 & 0.78  & 0.93          & 0.92          & 0.86 &  & 0.95          & 0.83 &  & 0.95    & \textbf{0.97} & 0.95 &  & 0.92          &  & 0.86     &  \\
Zeek \cite{Zeek}                           & 0.92          & 0.82 & 0.81  & 0.96          & 0.89          & 0.91 &  & 0.82          & 0.69 &  & 0.94    & \textbf{0.98} & 0.94 &  & 0.93          &  & 0.76     &  \\
Trex \cite{Trex}                           & 0.89          & 0.81 & 0.76  & \textbf{0.99} & 0.95          & 0.89 &  & 0.90          & 0.75 &  & 0.92    & \textbf{0.99} & 0.92 &  & 0.75          &  & 0.82     &  \\
HermesSim \cite{HermesSim}                 & \textbf{1.00} & 0.97 & 0.96  & \textbf{1.00} & \textbf{1.00} & 0.99 &  & \textbf{1.00} & 0.99 &  & 0.95    & 0.98          & 0.95 &  & \textbf{1.00} &  & 0.70     &  
\end{tblr}
}
\captionsetup{justification=centering, margin=0.5cm}
\caption*{\small The letter combinations denote how the function pairs were generated (e.g., XA: Cross-Architecture; XB: Cross-Bitness; XM: Arbitrary Architectures, Bitness,
Compiler, Compiler versions, and Optimization level, etc.). See Tables~\ref{tab:tasks} for more information. }
\end{table*}
\vspace{-0.5\baselineskip}

%% file: paper/04_evaluation.tex
\subsection{Model Selection}
\label{model selection}

\input{tables/d1results}

\begin{figure*}[!t]
    \centering
    \begin{subfigure}[b]{0.48\textwidth}
        \includegraphics[width=\linewidth]{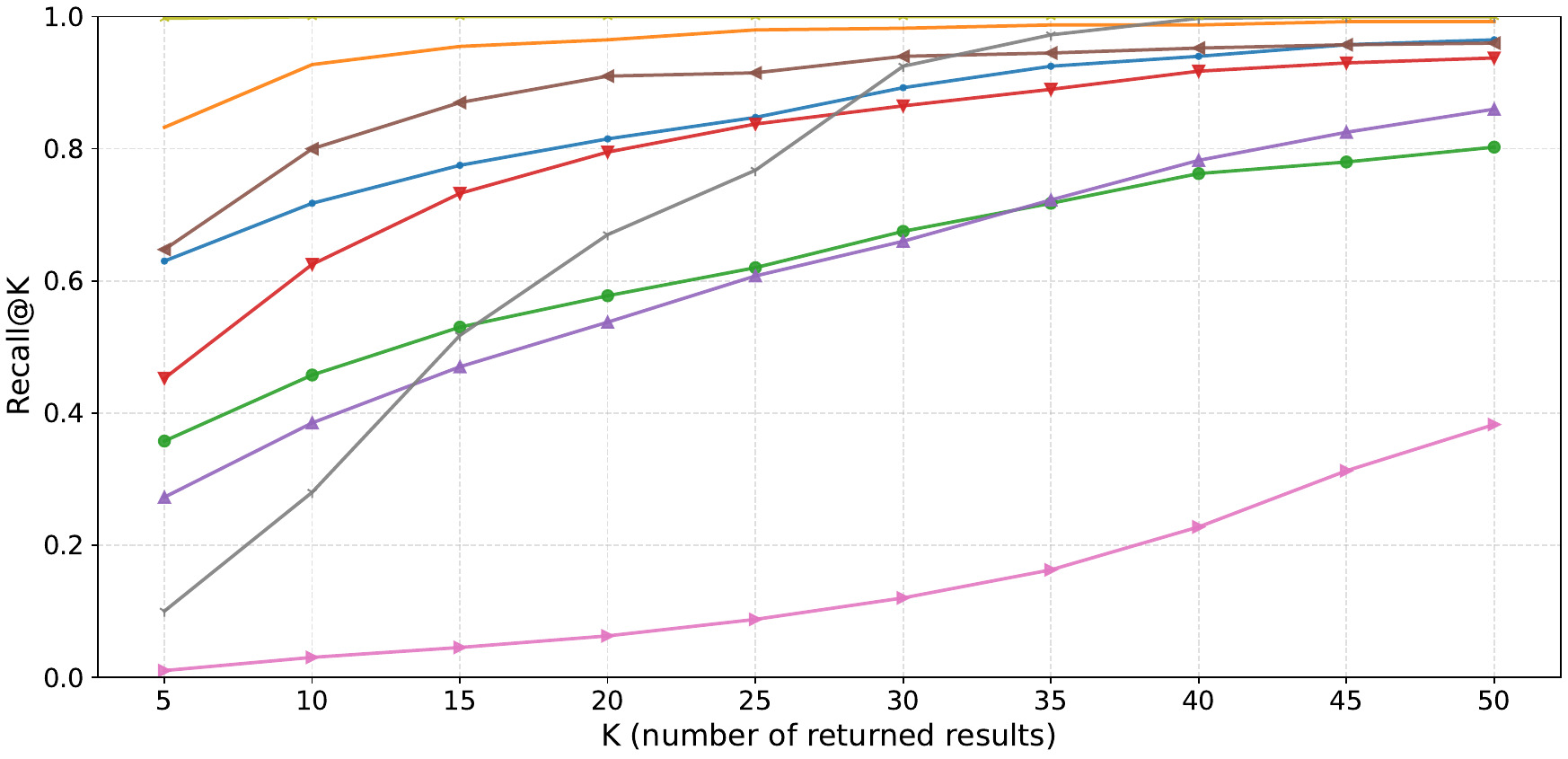}
        \caption{XA+XB}
    \end{subfigure}
    \hfill
    \begin{subfigure}[b]{0.48\textwidth}
        \includegraphics[width=\linewidth]{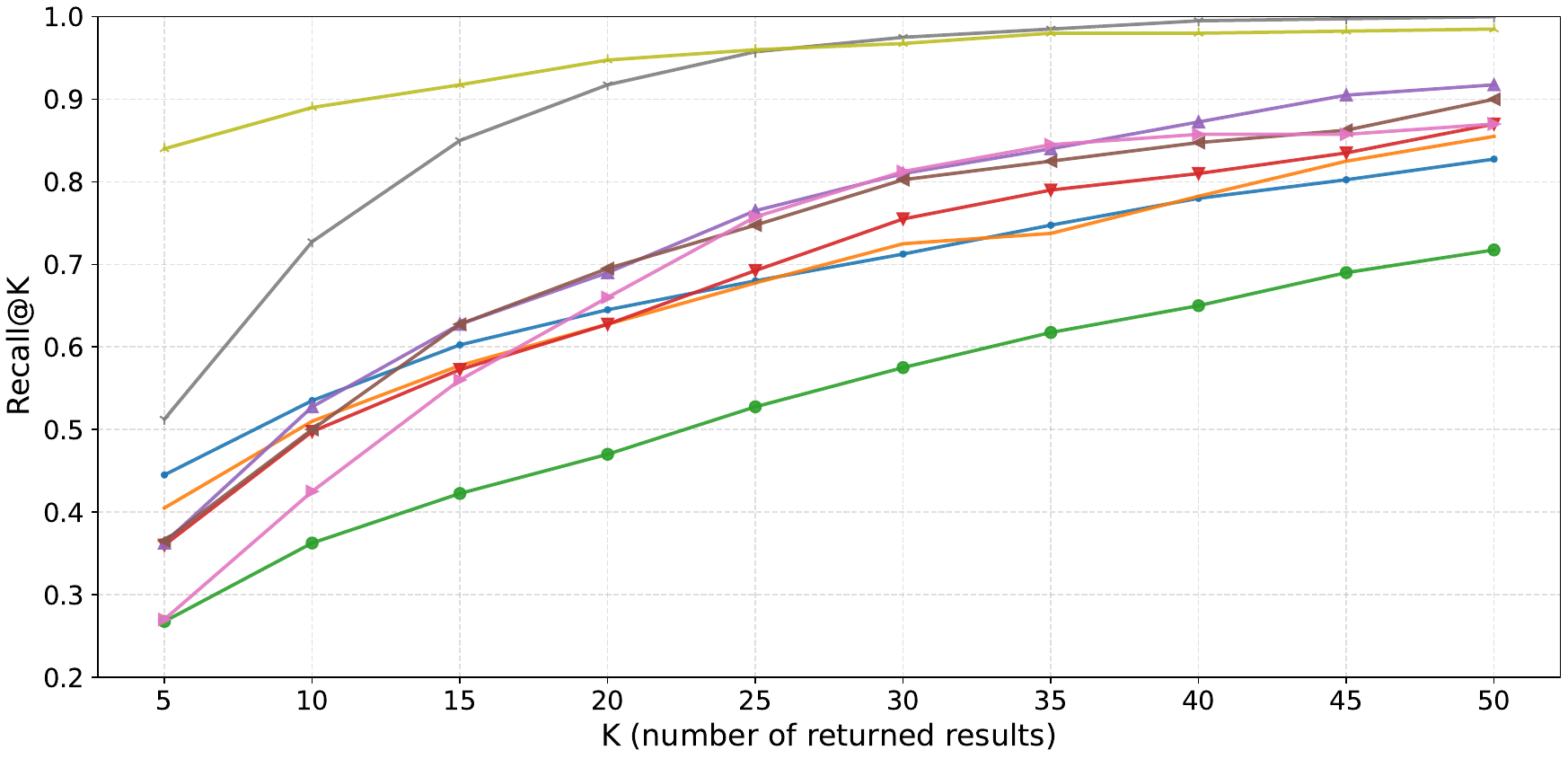}
        \caption{XC}
    \end{subfigure}
    
    \vspace{0.5em}
    
    \begin{subfigure}[b]{0.48\textwidth}
        \includegraphics[width=\linewidth]{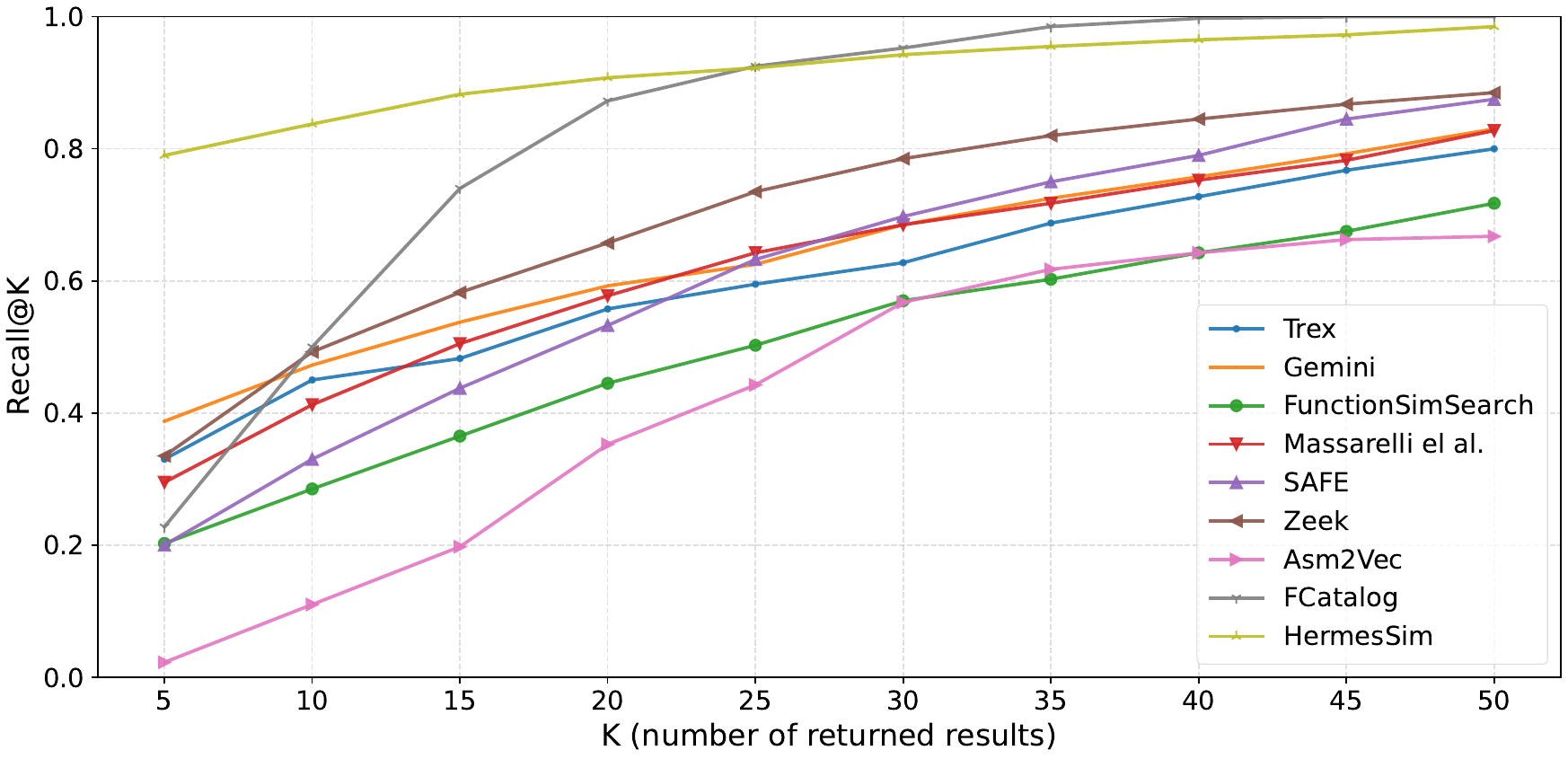}
        \caption{XC+XB}
    \end{subfigure}
    \hfill
    \begin{subfigure}[b]{0.48\textwidth}
        \includegraphics[width=\linewidth]{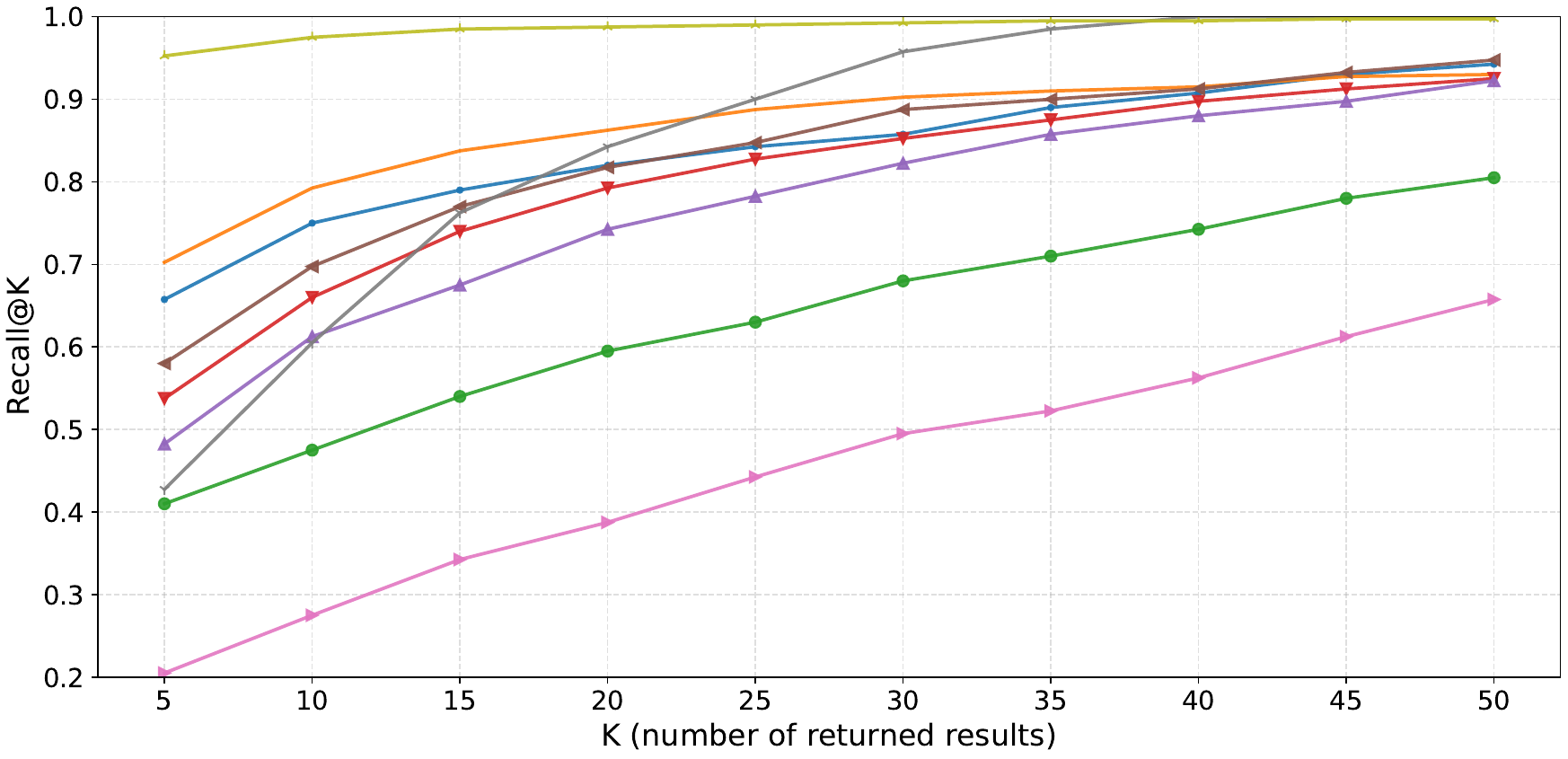}
        \caption{XM}
    \end{subfigure}

    \caption{Comparison of the recall at different K values for the Standard Dataset.}
    \label{fig:d1recall}
\end{figure*}

\begin{figure}[!b]
    \centering
    \begin{subfigure}[b]{0.48\textwidth}
        \includegraphics[width=\linewidth]{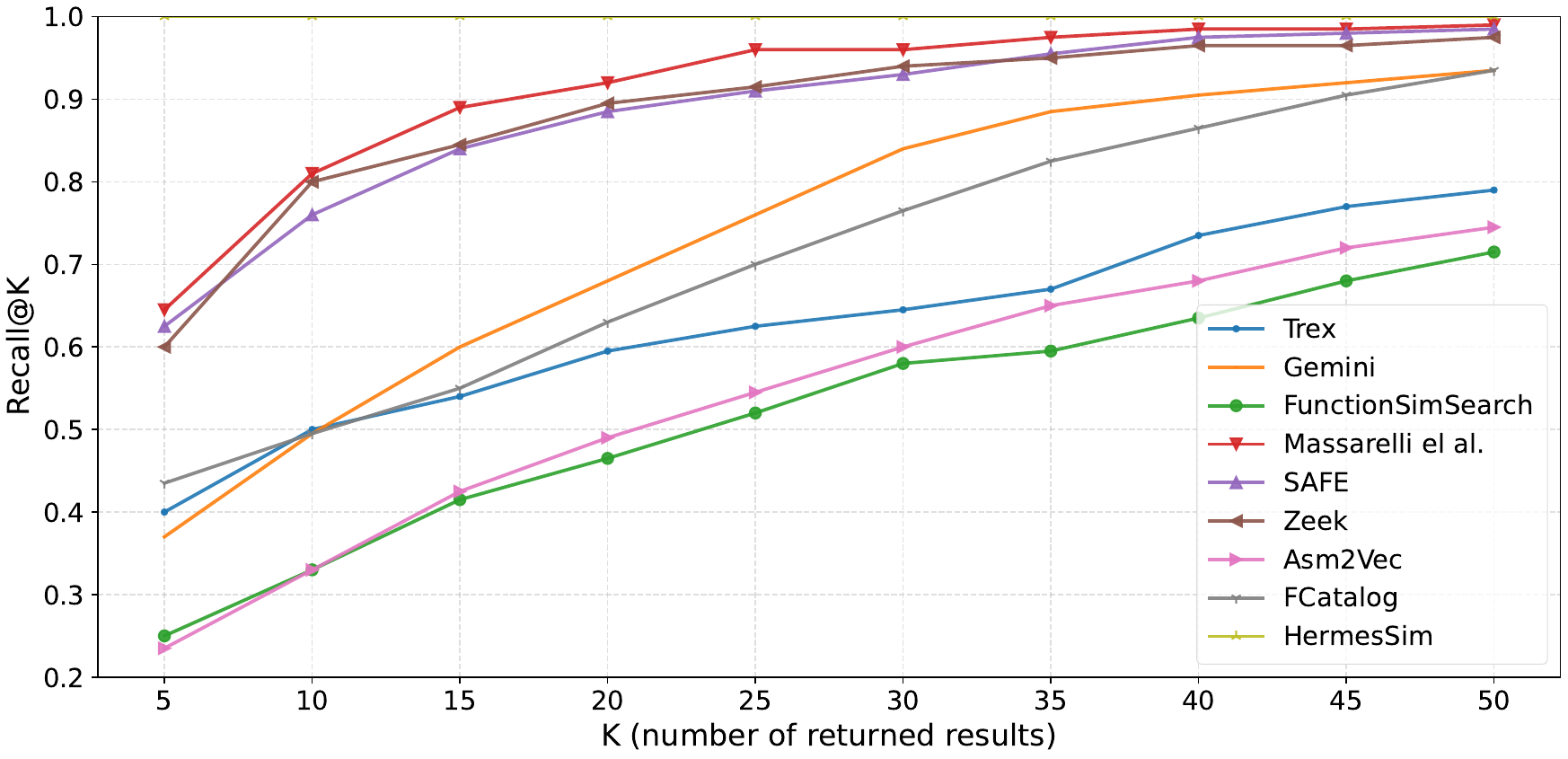}
        \caption{XOb for the Obfuscation Dataset.}
        \label{fig:d4recall}
    \end{subfigure}
    
    \vspace{0.8em}

    \begin{subfigure}[b]{0.48\textwidth}
        \includegraphics[width=\linewidth]{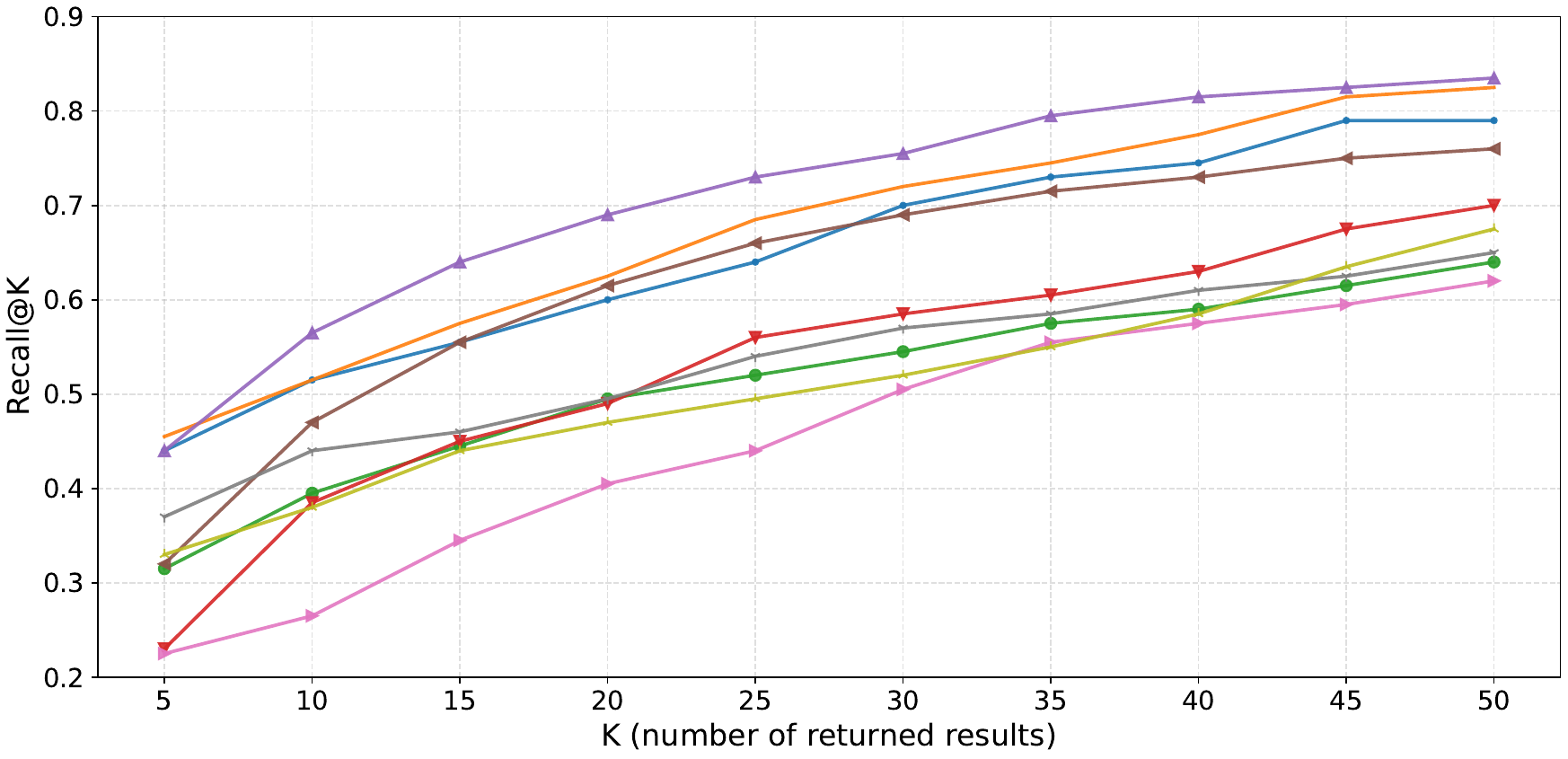}
        \caption{XS for the Semantic Dataset}
        \label{fig:d5recall}
    \end{subfigure}

    \caption{Recall graphs of Obfuscation and Semantic Dataset}
    \label{fig:d4d5recall}
\end{figure}

To ensure fair and scalable evaluation under diverse datasets, we restrict our selection to representative BFSD models with publicly available implementations and demonstrated robustness across heterogeneous inputs. As discussed in Section~\ref{Methods Systematization}, there are three main trends for solutions of the BFSD problem. Most of the models that we selected also fall into these three categories.


\paragraph{Model Selection.}
To evaluate the impact of dataset diversity across fundamentally different methodological assumptions in BFSD, we select representative approaches spanning three major paradigms: fuzzy hashing, graph-based learning, and NLP-inspired modeling. Our selection prioritizes publicly available and reproducible implementations that have demonstrated competitive performance in prior evaluations, ensuring fair and scalable experimentation under heterogeneous datasets.

\paragraph{Fuzzy Hashing.}
For fuzzy hashing, we include FCatalog \cite{FCatalog} and FunctionSimSearch \cite{functionsimsearch}, representing byte-level and CFG-aware locality-sensitive hashing strategies, respectively. These methods capture similarity through signature-based approximations rather than learned embeddings, serving as a contrast to neural approaches.

\paragraph{Graph-Based Methods.}
For graph-based learning, we select Gemini \cite{gemini} and HermesSim \cite{HermesSim}, two representative GNN-based models that operate on structured control-flow representations. While Gemini builds upon attributed control-flow graphs, HermesSim emphasizes semantic enrichment through alternative graph representations. We exclude classical graph-matching tools such as BinDiff \cite{bindiff}, as our study focuses on learning-based BFSD methods.

\paragraph{NLP-Based Methods.}
For NLP-inspired approaches, we include Asm2Vec \cite{Asm2Vec}, SAFE \cite{SAFE}, and Trex \cite{Trex}, covering three distinct modeling paradigms: token embedding (word2vec-style), sequence modeling (seq2seq-style), and Transformer-based representations. These models treat binary code as structured sequences rather than graphs, enabling comparison across representation assumptions.

\paragraph{Hybrid Approaches.}
We additionally include Massarelli et al. \cite{Massarelli} and Zeek \cite{Zeek} as hybrid approaches that integrate graph analysis with neural semantic modeling. Their inclusion allows us to assess how dataset diversity affects methods that combine structural and token-level representations.

\subsection{Implementation}
\label{Implementation}

For fairness, we used IDA Pro 7.6 \cite{IDAPro} for binary analysis for all models. Although IDA Pro is not freely available, it remains the industry-standard disassembler with the most accurate function-boundary recovery and control-flow graph reconstruction for binaries. We use IDA Pro to minimize errors in function extraction, and we export all intermediate JSON/CFG artifacts so that any researcher can re-run the exact same benchmarking pipeline with a free tool (e.g., Ghidra \cite{ghidra} and Radare2 \cite{radare2}) using our provided exports. 

Some models require feature extraction; for example, graph-based methods require ACFG. We use IDA Pro plugins, Capstone \cite{capstone}, and NetworkX \cite{networkx} in Python scripts to cater to such needs. We ran HermesSim with the open source project provided by its authors, and the other models are from the repository created by the authors of \cite{usenix22}.

All experiments are run on a computer with Linux Ubuntu 22.04, two Intel(R) Xeon(R) Silver 4310 CPU @ 2.10GHz, 256GB DDR4 RAM, 17TB SSD, and one NVIDIA A30 GPU.

\input{tables/d2results}
\subsection{Experiment Design}
\label{Experiment Design}
\input{tables/d3results}

We evaluated the \modelNumber models on the 5 datasets that we created for the evaluation section. For each dataset, we divide it into training, validation, and testing sets. When we evaluate a model on a dataset, we first do feature extraction for the subsets; then we train the model with the training and validation set of the dataset before running inference on the testing set (Except Trex \cite{Trex}, as it is pretrained.). The ground truth is determined by the function name deciphered in the feature extraction process. If two functions are from binaries of the same source code and have the same function name, they're considered a positive match. We filter out duplicate function pairs where the content hashes of the two functions are identical, as well as functions that contain less than 5 basic blocks.

For the results presentation, we adopted multiple measures to reflect the performance and efficiency of the evaluated models on the datasets. First, we use the area under curve (AUC) of the receiver operating characteristic (ROC) curve as the most straightforward and significant indicator of the performance of a model, because AUC denotes the overall classification performance across all possible thresholds. 

\[
\text{Recall} = \frac{\text{True Positives}}{\text{True Positives} + \text{False Negatives}}
\]

We also include recall and mean reciprocal rank (MRR) as complimentary measures for the models' performance. As shown above, recall is defined as the rate of true positives among all positives. And pecall@1 will reflect how likely the correct match appears in the top 1 result. MRR denotes on average, how early the first correct result shows up in the ranking. As shown below, MRR@10 is the same as MRR, but only counts the reciprocal rank if the correct result is in the top 10. If there's no correct ones in the top 10 results, the score will be 0.

\[
\text{MRR} = \frac{1}{N} \sum_{i=1}^{N} \frac{1}{\text{rank}_i}
\]

\[
\text{MRR@10} = \frac{1}{N} \sum_{i=1}^{N}
\begin{cases}
\frac{1}{\text{rank}_i}, & \text{if } \text{rank}_i \leq 10 \\
0, & \text{otherwise}
\end{cases}
\]

For efficiency, we present the feature extraction time and the inference time for the testing set of each model running on every dataset. Because these two running times are usually large, we also calculate $Tot 100$, which is the combined previous two running times that the model takes per 100 functions processed, on average. 

\subsection{Inter‐Dataset Comparison}
\label{Inter‐Dataset Comparison}
Table~\ref{tab:AUC_overview} summarizes each model's best AUC on one representative task per dataset. Below, we highlight key trends using those values.

\paragraph{Standard Dataset.} This dataset primarily evaluates low-level differences in our taxonomy (e.g., architecture, compiler, optimization). On Standard, HermesSim achieves perfect AUC ($1.00$) on both XA and XV tasks, with Gemini close behind at $0.96$ (XA) and $0.98$ (XV).  SAFE and Zeek yield AUCs in the low‐to‐mid $0.90$ range (e.g., SAFE: $0.78$–$0.93$; Zeek: $0.92$–$0.96$), while FCatalog and Asm2Vec remain further down ($0.44$–$0.99$ depending on task).  FunctionSimSearch sits around $0.65$–$0.81$ and Trex ranges from $0.76$–$0.99$.  In short, most models peak on Standard, with the GNN‐based methods (HermesSim, Gemini) clearly leading.

\paragraph{Firmware Dataset.} Firmware also corresponds primarily to low-level variation, but introduces real-world compilation diversity and architectural unpredictability. On Firmware, HermesSim again tops the table ($1.00$ and $0.99$). Clever token‐plus‐graph models like SAFE achieve $0.95$ on XV, while Massarelli et al.\ reaches $0.91$ on XV and $0.80$ on XO.  Gemini drops slightly to $0.96$ on XV. FCatalog remains strong at $0.99$–$0.90$, but Zeek falls to $0.82$–$0.69$. Lower‐complexity methods such as Asm2Vec decline further ($0.62$–$0.57$), and FunctionSimSearch sits near $0.74$–$0.61$. Overall, firmware binaries introduce enough architectural variation that only the top graph‐based and fuzzy hashing methods maintain AUC above $0.90$.


\paragraph{Malware Dataset.}
Malware primarily reflects low-level differences in our taxonomy, but in a less controlled and more naturally occurring setting than Standard. Rather than being generated through systematic recompilation, these binaries are collected from the wild and exhibit real-world architectural and compilation diversity. Interestingly, most models perform at or near their best on Malware (XC/XV/XO tasks). HermesSim records $0.95$–$0.98$, Gemini achieves $0.89$–$0.99$, and SAFE/Zeek both lie at $0.95$–$0.98$. FCatalog ranges from $0.83$–$0.99$, and Asm2Vec even reaches $0.92$–$0.99$, while Massarelli et al.\ and Trex likewise exceed $0.90$–$0.99$.

\paragraph{Obfuscation Dataset.} This dataset isolates mid-level differences in our taxonomy by applying semantics-preserving obfuscation transformations. When obfuscation techniques are applied, AUC values drop for nearly all models compared to other datasets. HermesSim still stands out with a perfect score of $1.00$. The next tier includes Zeek ($0.93$), SAFE ($0.92$), and Massarelli et al.\ ($0.92$). Gemini follows at $0.84$, while FCatalog ($0.78$) and Trex ($0.75$) show weaker robustness. Asm2Vec ($0.67$) and FunctionSimSearch ($0.64$) perform the worst under obfuscation. The compressed AUC range (from $0.64$ up to $1.00$) highlights both the increased difficulty of this setting and the clear advantage of HermesSim over prior models.

\paragraph{Semantic Dataset.} This directly tests the third level of our taxonomy. On the Semantic Dataset, SAFE achieves the highest AUC ($0.86$), followed closely by Gemini ($0.83$) and Trex ($0.82$). Zeek and Massarelli et al.\ reach $0.76$ and $0.73$, respectively, while FCatalog and HermesSim both report $0.70$. FunctionSimSearch attains $0.68$, and Asm2Vec yields the lowest AUC at $0.53$. In terms of Recall@1 and MRR@10, Trex leads with $0.34$ and $0.33$, respectively, with Gemini close behind at $0.33$ and $0.30$. SAFE performs comparably on Recall@1 ($0.30$) and MRR@10 ($0.28$), whereas the other models remain below $0.27$. Overall, the AUC range ($0.53$–$0.86$) is narrow and relatively low compared to other datasets, underscoring the challenge of capturing semantically equivalent functions across diverse source implementations.


\paragraph{Summary.}
Taken together, the five datasets illustrate that BFSD robustness is strongly taxonomy-dependent. Under low-level differences (Standard and Malware), several models achieve near-perfect AUC (e.g., HermesSim at $1.00$), with large inter-model gaps. Firmware values generally decline by $0.05$–$0.10$ relative to Standard, while Obfuscation and Semantic datasets push most AUCs below $0.90$. The most substantial degradation occurs under high-level semantic variation, where AUCs fall to $0.53$–$0.86$, representing up to a $30\%$ drop compared to Standard. No single model dominates across all three levels of variation, confirming that robustness cannot be treated as a monolithic property.

\subsection{Inter-Model Comparison}
\label{Inter-Model Comparison}
\input{tables/d4results}
\input{tables/d5results}
In this section, we present detailed performance and efficiency scores on each dataset (see Tables~\ref{tab:d1comparison}–\ref{tab:d5comparison}), where each table corresponds respectively to the 5 datasets we have, with each row shows the performance data of a selected model, and each column represents a performance metric. Because the purpose of these tables is to compare the performance of different models on the same metric with the same dataset, the highest performance score is highlighted in bold for each metric. 

\subsubsection{Performance}

\paragraph{Fuzzy Hashing.} Surprisingly, FCatalog consistently outperforms FunctionSimSearch and even rivals several modern ML-based approaches across multiple datasets. This result is notable given that FCatalog contains no learned components and relies purely on carefully engineered fuzzy hashing features. In particular, FCatalog \cite{FCatalog} outperforms many models on the Firmware Dataset (Table~\ref{tab:d2comparison}), and remains competitive in several tasks within the Standard and Malware datasets. It also achieves respectable Recall and MRR@10 on the Semantic Dataset, though not among the top performers.

As shown in Figure~\ref{fig:d1recall}, FCatalog \cite{FCatalog} tends to start low on Recall but quickly rises as $K$ increases. By contrast, FunctionSimSearch \cite{functionsimsearch} consistently shows the weakest Recall across all $K$ values. In Figure~\ref{fig:d4d5recall}, although FCatalog is not the leader, it maintains mid-tier Recall performance on the Obfuscation and Semantic datasets, while FunctionSimSearch remains near the bottom.

In conclusion, FCatalog consistently outperforms FunctionSimSearch across all five datasets, demonstrating that even without deep learning, well-designed fuzzy hashing techniques remain competitive, especially as $K$ grows. FunctionSimSearch, by contrast, struggles to keep pace once $K$ exceeds small values, suggesting that its similarity metrics lack sufficient discrimination in more complex settings. Overall, our results show that fuzzy hashing remains a useful baseline for function similarity detection — efficient and interpretable, though no longer state-of-the-art.

\paragraph{Graph Based Methods.} Graph representations excel at modeling relationships between instructions, both control flow and data flow, enabling them to capture higher-level semantic features that token-based or sequence-based methods cannot. In our experiments, Gemini \cite{gemini} and HermesSim \cite{HermesSim} consistently outperformed other categories, confirming that structural information is critical for accurate function similarity detection. On the Obfuscation Dataset (Table~\ref{tab:d4comparison}), HermesSim achieved perfect performance ($\text{AUC} = 1.00$, $\text{MRR@10} = 1.00$, $\text{Recall@1} = 1.00$), clearly surpassing all other models and demonstrating that its Semantics-Oriented Graph (SOG) representation is especially robust to heavy code transformations. By contrast, non-graph methods often struggled to distinguish between inlined or reordered basic blocks once obfuscation was introduced. On the Semantic Dataset, however, HermesSim dropped to $0.70$ AUC, while Gemini reached $0.83$, suggesting that Gemini’s attributed control-flow embeddings are better tuned for purely semantic variations.

From Figure~\ref{fig:d1recall}, Gemini leads in all four tasks of the Standard Dataset, achieving the highest Recall when $K$ is small in three out of four cases. This strong initial performance suggests that Gemini’s GNN over Attributed Control-Flow Graphs correctly prioritizes the most critical structural similarities before deeper semantic cues become necessary. On the Semantic Dataset (Figure~\ref{fig:d5recall}), HermesSim’s intra-instruction tokenization helps narrow the gap, though Gemini still rises faster in Recall as $K$ increases, indicating a more balanced trade-off between precision and coverage. Finally, in Figure~\ref{fig:d4recall}, HermesSim consistently outperforms every other model at every $K$ value, confirming that its SOG representation, by filtering out irrelevant tokens and normalizing instruction semantics, produces remarkably stable high Recall regardless of sample size. These results demonstrate that graph-based approaches not only achieve the top aggregate scores but also provide greater resilience when binaries undergo aggressive optimization, obfuscation, or semantic modification.

\paragraph{NLP Based Methods.} NLP-based approaches treat disassembled instructions as token sequences, learning embeddings that capture lexical and syntactic patterns but not necessarily detailed control-flow structure. In our evaluation, SAFE \cite{SAFE}, Massarelli et al.\ \cite{Massarelli}, and Zeek \cite{Zeek} form the strongest tier of NLP-based models. SAFE achieves consistently high AUC, including $0.92$ on the Obfuscation Dataset, thanks to its hybrid token embeddings and lightweight control-flow summarization. Massarelli et al.\ combine attributed control-flow graphs with token embeddings in a joint neural model, reaching top scores on the Standard ($0.97$) and Malware ($0.98$) datasets. Zeek, which lifts functions into an intermediate representation and builds data-flow graphs, achieves $0.93$ AUC under obfuscation, showing that structural hints even in NLP-heavy models can substantially improve robustness.

Trex and Asm2Vec perform notably worse. Trex lags on both Obfuscation ($0.75$) and Semantic ($0.69$), while Asm2Vec, one of the earliest purely NLP-driven methods, generally records the lowest AUC, Recall, and MRR@10 across most datasets (e.g., $0.67$ on Obfuscation, $0.63$ on Semantic). The only exception is the Malware Dataset, where Asm2Vec avoids last place, likely because malware families share idiosyncratic instruction patterns that its token embeddings can partially capture.

Overall, NLP-based methods remain interpretable and computationally efficient baselines, but they are outperformed by graph-based approaches such as Gemini \cite{gemini} and HermesSim \cite{HermesSim}, especially under heavy obfuscation or cross-architecture/compiler variation. Their competitiveness improves when combined with structural features, as shown by SAFE, Zeek, and Massarelli et al., but purely token-based embeddings struggle to scale across more complex transformation scenarios.

\subsubsection{Efficiency}

\begin{figure}[!t]
  \centering
  \begin{subfigure}[b]{0.45\textwidth}
    \centering
    \includegraphics[width=\linewidth]{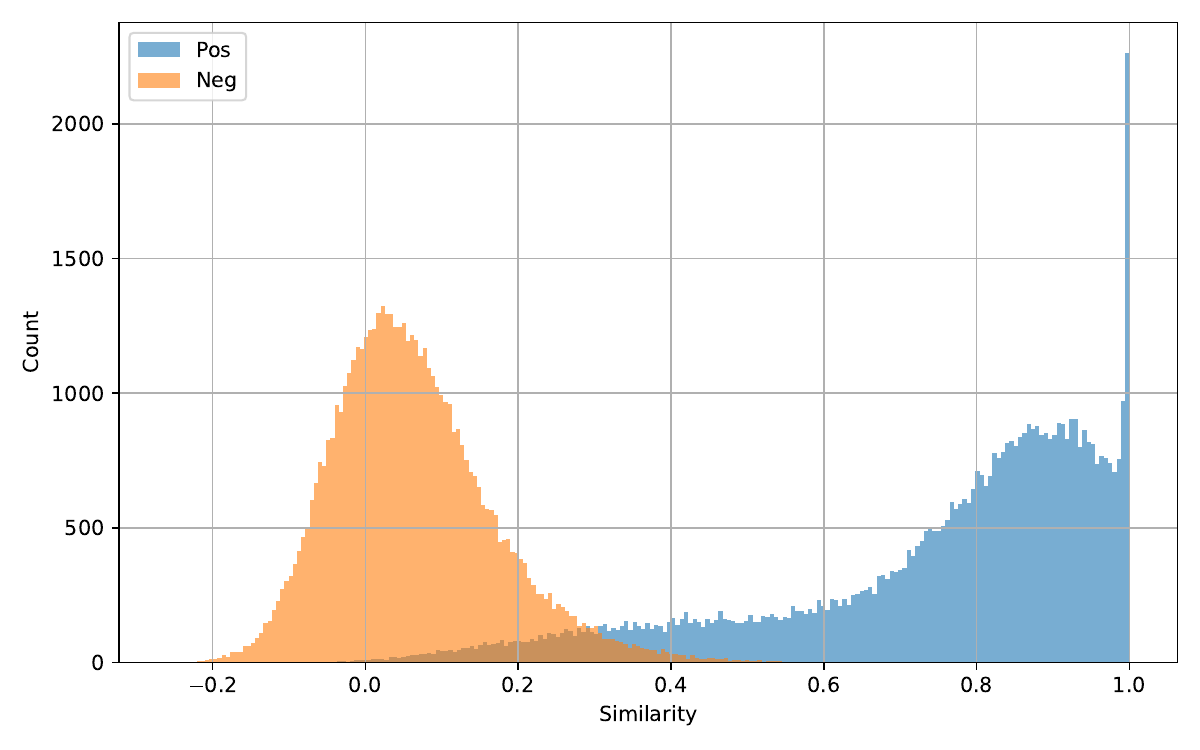}
    \caption{HermesSim on Standard Dataset}
    \label{fig:d1h}
  \end{subfigure}


  \begin{subfigure}[b]{0.45\textwidth}
    \centering
    \includegraphics[width=\linewidth]{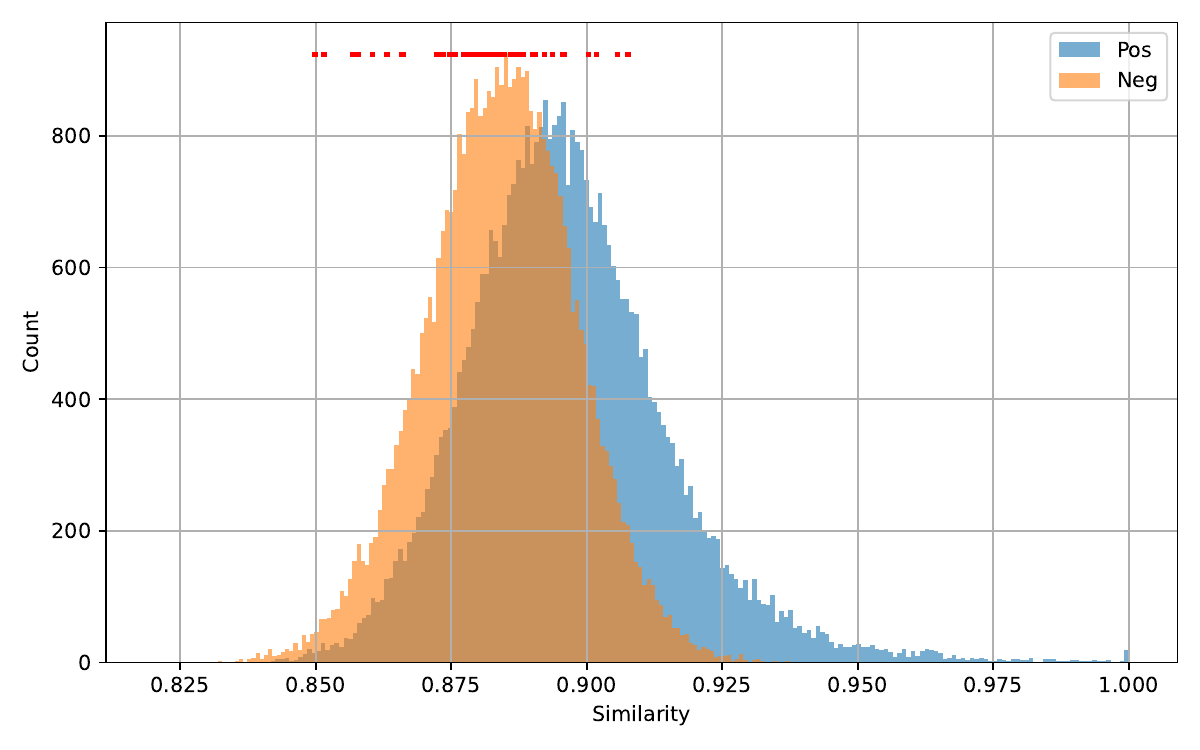}
    \caption{HermesSim on Semantic Dataset}
    \label{fig:d5h}
  \end{subfigure}


  \begin{subfigure}[b]{0.45\textwidth}
    \centering
    \includegraphics[width=\linewidth]{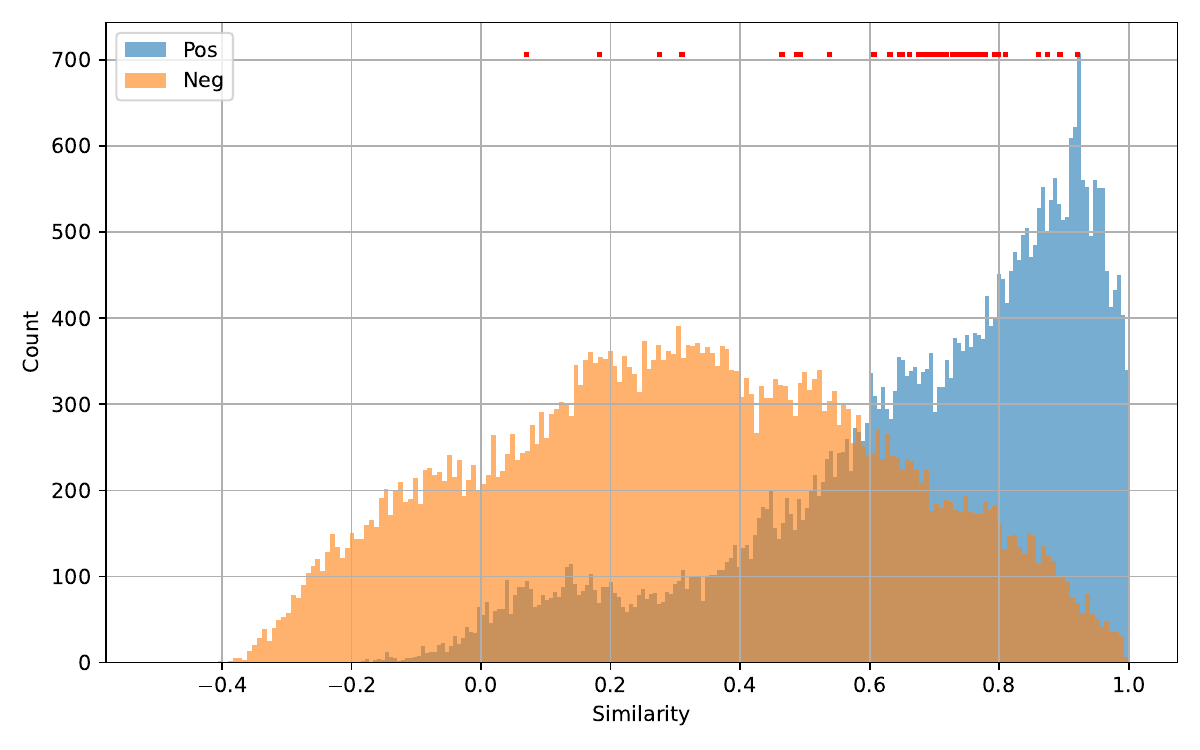}
    \caption{Trex on Semantic Dataset}
    \label{fig:d5t}
  \end{subfigure}

  \caption{A Comparison of the Similarity Score Distributions.}
  \label{fig:sim_comparison}
\end{figure}

\paragraph{Fuzzy Hashing.} FCatalog \cite{FCatalog} remains highly efficient since it requires only a single feature-extraction pass, averaging just a few seconds per 100 functions on most datasets (e.g., $3.34$\,s on Firmware, $6.33$\,s on Obfuscation). FunctionSimSearch \cite{functionsimsearch} is only marginally slower, typically about $1.1$--$1.3\times$ FCatalog's runtime (e.g., $0.37$\,s vs.\ $3.34$\,s on Firmware, $7.26$\,s vs.\ $6.33$\,s on Obfuscation). By contrast, graph-based methods such as Gemini and HermesSim incur far higher costs, often hundreds of seconds per 100 functions on larger datasets. Thus, while structural models dominate accuracy, classical approaches like FCatalog and FunctionSimSearch remain the most efficient.

\paragraph{Graph Based Methods.} Graph techniques incur substantially higher runtime costs due to structural analysis. On the Malware Dataset (Table~\ref{tab:d3comparison}), HermesSim \cite{HermesSim} requires $326.17$\,s per 100 functions, ranking among the slowest models despite achieving near-perfect $\mathrm{AUC}=0.95$. By contrast, Gemini \cite{gemini} processes the same split in only $4.73$\,s per 100 while reaching the same $\mathrm{AUC}=0.95$. On the Obfuscation Dataset (Table~\ref{tab:d4comparison}), HermesSim again dominates accuracy with $\mathrm{AUC}=1.00$, $\mathrm{MRR@10}=1.00$, and $\mathrm{Recall@1}=1.00$, but takes $33.22$\,s per 100 functions. Gemini, meanwhile, runs nearly $3\times$ faster at $12.76$\,s but lags behind in accuracy ($\mathrm{AUC}=0.84$, $\mathrm{Recall@1}=0.16$). Thus, HermesSim sets the accuracy ceiling, while Gemini offers a faster, more scalable compromise for large--scale or latency-sensitive scenarios.

\paragraph{NLP Based Methods.} Token-only or hybrid methods run faster than full GNN analyzers but still vary widely in cost. On the Standard Dataset (Table~\ref{tab:d1comparison}), Asm2Vec \cite{Asm2Vec} requires only $1.15$\,s per 100 functions with $\mathrm{AUC}=0.44\text{--}0.81$, while SAFE \cite{SAFE} is even faster at $0.13$\,s with $\mathrm{AUC}\ge0.76$. Zeek \cite{Zeek}, which combines data-flow graphs with token embeddings, is slower on Standard ($30.33$\,s, $\mathrm{AUC}=0.81\text{--}0.96$) but more efficient on Firmware and Obfuscation ($2\text{--}3$\,s, $\mathrm{AUC}\approx0.82\text{--}0.93$). Massarelli et al.\ \cite{Massarelli} occupy the middle ground, costing $1.84$\,s on Standard and $17.26$\,s on Malware with $\mathrm{AUC}\approx0.90$. Trex \cite{Trex} is the clear runtime outlier: $11.60$\,s on Standard and over $1800$\,s on Semantic despite only moderate performance ($\mathrm{AUC}=0.82$, $\mathrm{Recall@1}=0.34$). In summary, Asm2Vec and SAFE sit at the high-efficiency end, Massarelli et al.\ and Zeek are mid-range, and Trex remains the slowest NLP-based model.

\paragraph{Summary.} In general, classical fuzzy-hashing methods provide the strongest efficiency, NLP-based approaches balance runtime and semantic capacity, and graph-based techniques achieve the highest accuracy at substantially higher computational cost. These results highlight a consistent trade-off between scalability and structural modeling power across BFSD paradigms.

\subsection{Case Study}
\label{Case Study}

In the previous sections, we observed an intriguing phenomenon on the Semantic Dataset. HermesSim \cite{HermesSim} achieves near-perfect AUC on most datasets, yet its performance drops sharply on the Semantic Dataset. Notably, HermesSim attributes its success to the proposed Semantic Oriented Graph, which is designed to emphasize semantic properties of binary functions. To better understand this discrepancy, we conduct a detailed case study.

\paragraph{Backgrounds.} To help better understanding the performance plummet, we analyzed the distribution of the similarity scores of both the positive and negative testing function pairs on the Standard Dataset and the Semantic Dataset, as shown in Figure~\ref{fig:sim_comparison}. To better compare with and study the performance of HermesSim on the Semantic Dataset, we also draw the HermesSim on Standard Dataset and Trex on Semantic Dataset in contrast.

As we can see, the shape of the similarity score distribution of HermesSim on Semantic Dataset is very different from the other two, where the positive and negative function pairs have separate peaks that are far from each other whereas the HermesSim on Semantic Dataset peaks are close and even overlapping for most part.





\begin{figure}[!tb]
  \centering
  \begin{subfigure}[b]{0.45\textwidth}
    \centering
    \includegraphics[width=\linewidth]{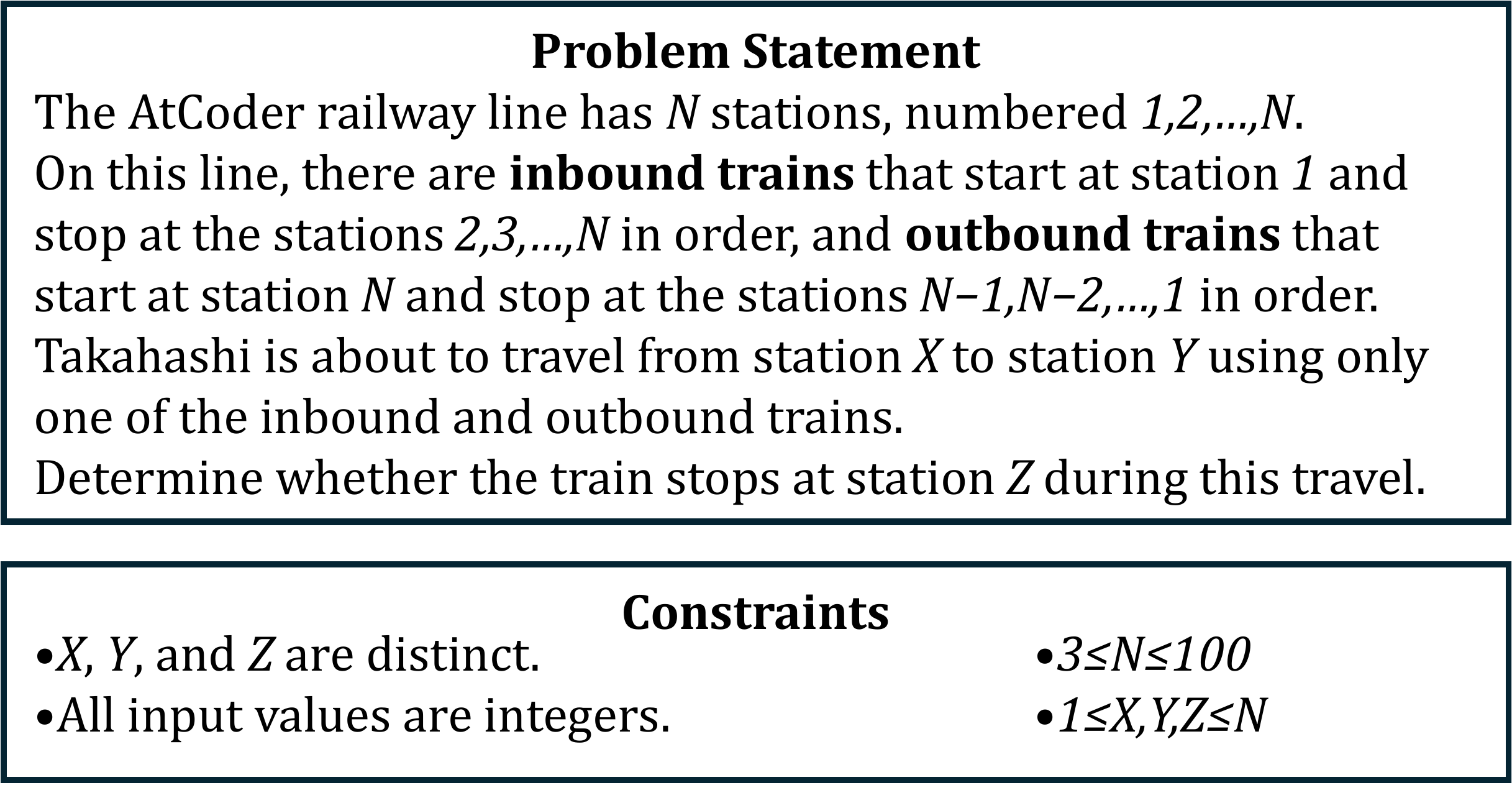}
    \caption{Problem Description}
    \label{fig:case_problem}
  \end{subfigure}

  \vspace{0.3cm} 

  \begin{subfigure}[t]{0.48\linewidth} 
    \centering
    \begin{lstlisting}
int n, x, y, z, cur, f=0;
scanf("%d%d%d%d",&n,&x,&y,&z);
cur = x;
while(1) {
    if (cur == z) f = 1;
    if (cur == y) break;
    if (x < y) cur++;
    else cur--;
}
if (f) puts("Yes");
else puts("No");
return 0;
    \end{lstlisting}
    \caption{Function No.1}
    \label{fig:sol1}
  \end{subfigure}
  \hfill
  \begin{subfigure}[t]{0.48\linewidth}
    \centering
    \begin{lstlisting}
int n, x, y, z;

scanf("%d%d%d%d",&n,&x,&y,&z);

if ((x < z && z < y) || 
    (y < z && z < x)) {
    puts("Yes");
} else {
    puts("No");
}

return 0;
    \end{lstlisting}
    \caption{Function No.2}
    \label{fig:sol2}
  \end{subfigure}

  \caption{Case Study Problem and Functions.}
  \label{fig:case_study}
\end{figure}

\paragraph{The Case.} We want to know why the Semantic Dataset is hard for HermesSim to solve, so we listed 5 positive function pairs with lowest similarity scores and the 5 negative pairs with the highest, as these are the most wrong cases. And as we peruse through the function names, we were surprised to find out that one function was in all 5 of the lowest-scored positive pairs, indicating that this function is special.

Recall how the Semantic Dataset was generated, each function is actually the main function of a submission to easy-level problem in a coding contest. And being a positive function pair with low similarity score means that there are two functions that solve the same simple problem, but HermesSim is not able to tell that these two are similar in semantics.

To help understanding this case, we present the lowest scored positive function pair in Figure~\ref{fig:case_study}, where the problem description is included and Function No.1 is the function that was causing the problem for HermesSim. As we can see, this is a fairly easy problem, but the solutions here use different methods to solve it. Function No.1 uses a more preliminary simulation-like method to check if the train passes the station while Function No.2, and most of the other solutions for this problem, uses a more mathematical and efficient method which checks some conditions for the numbers directly.

Although the semantic dataset contains many problems more difficult than this one, the failure of the selected model on this function pair shows that semantic differences are not necessarily positively correlated with the inherent complexity of the function itself.

Since Trex \cite{Trex} is the model that achieves top performance on the Semantic Dataset, we included the similarity score distribution of it in Figure~\ref{fig:sim_comparison}. And we used red dots to mark the similarity score of all the positive function pairs that involves Function No.1 in Figure~\ref{fig:d5h} and Figure~\ref{fig:d5t}. We can see that HermesSim handle all of these positive function pairs rather poorly while Trex was able to correctly mark most of them with higher similarity scores.

\begin{figure}[!tb]
  \centering
  \begin{subfigure}[b]{0.22\textwidth}
    \centering
    \includegraphics[width=\linewidth]{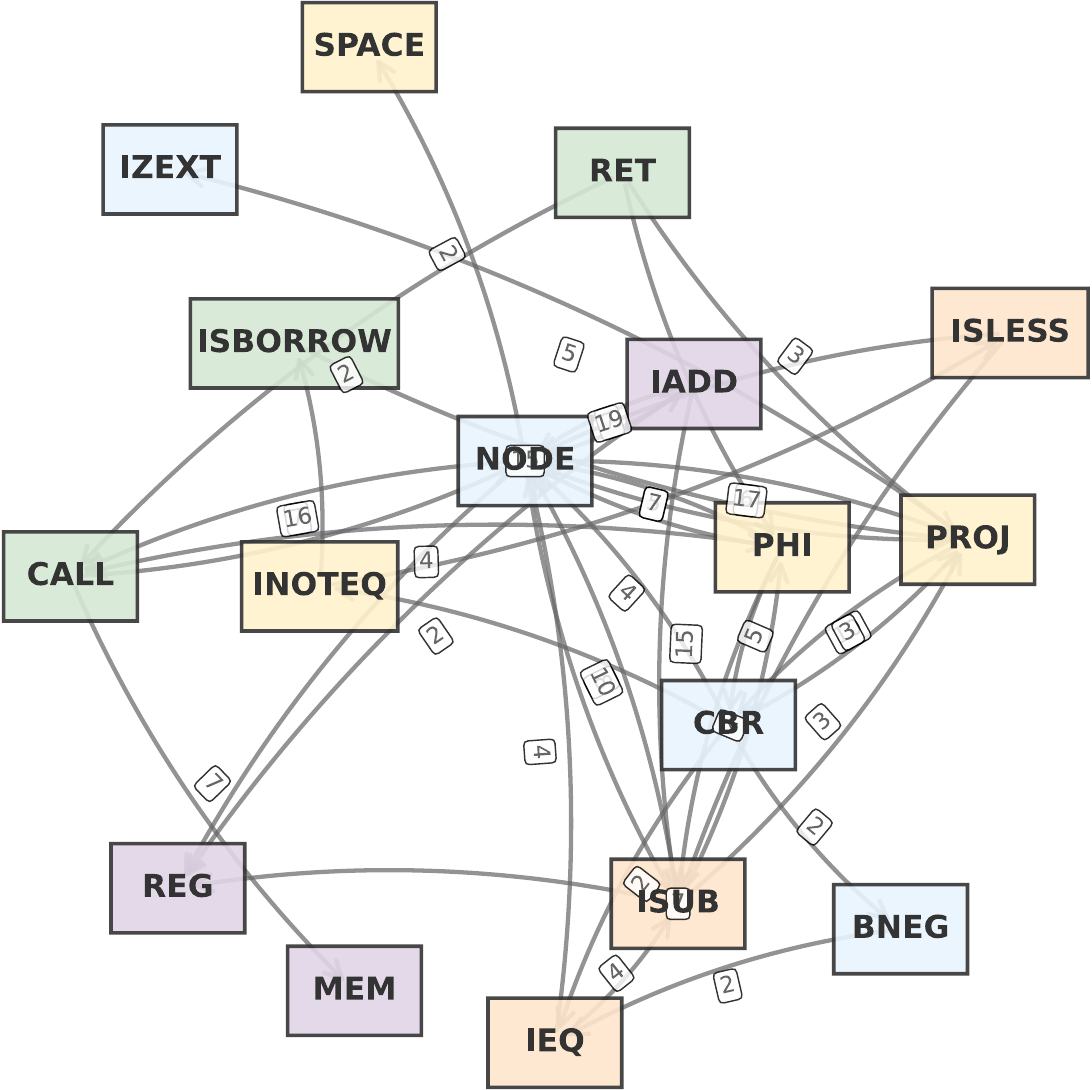}
    \caption{SOG of Function No.1}
    \label{fig:sol1sog}
  \end{subfigure}
  \hfill
  \begin{subfigure}[b]{0.22\textwidth}
    \centering
    \includegraphics[width=\linewidth]{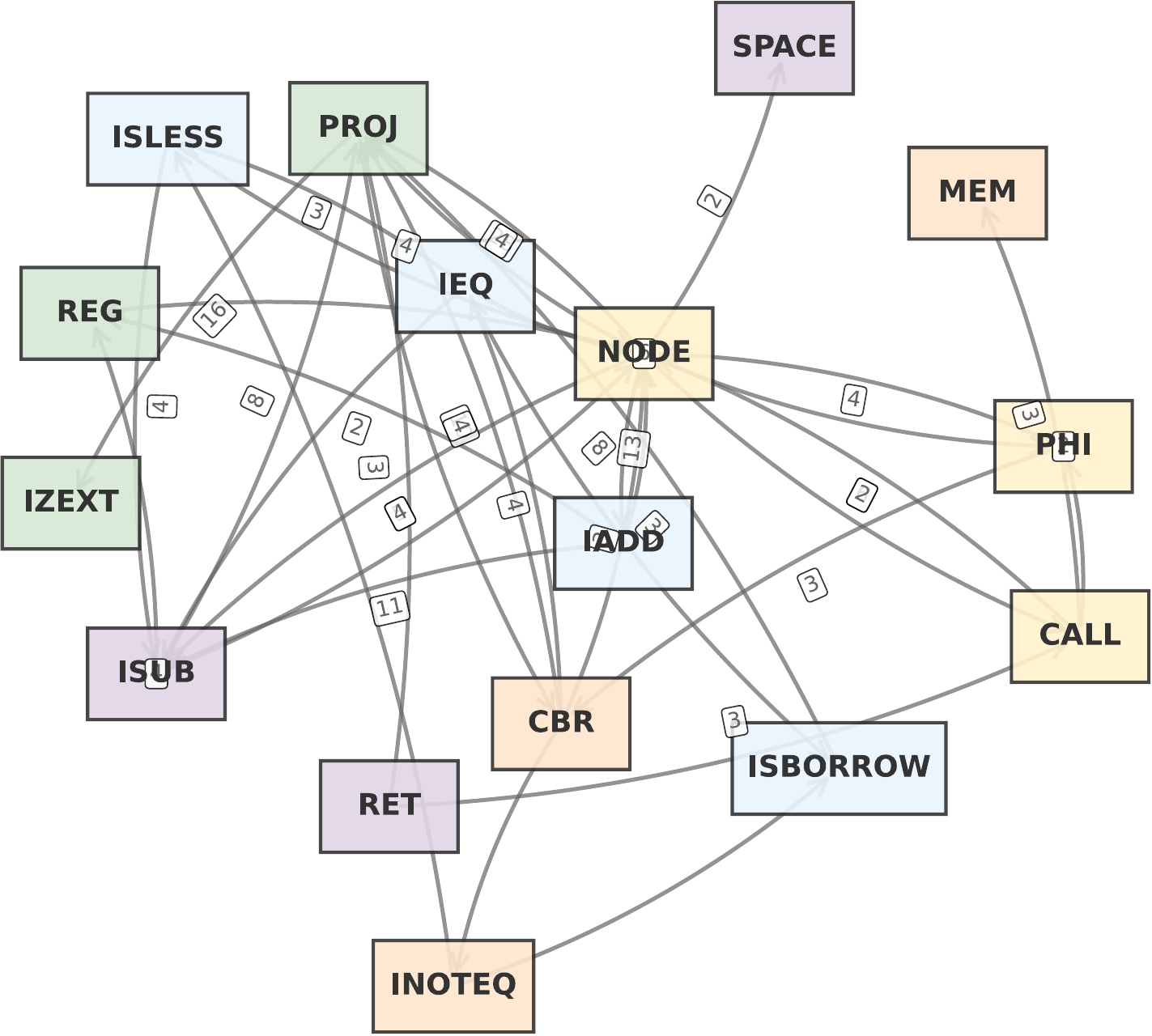}
    \caption{SOG of Function No.2}
    \label{fig:sol2sog}
  \end{subfigure}

  \caption{Comparison of the SOGs of the Two Functions.}
  \label{fig:sogs}
\end{figure}

\paragraph{Analysis.} To find out the cause of the different performance of these two models, we need to go back to their mechanisms. The core technique that HermesSim utilizes is the SOG, which can grasp and condense the semantic meanings of the functions studied, as claimed by its authors. We drew the SOGs created for Function No.1 and No.2 by the HermesSim preprocessing script and place them in contrast to each other in Figure~\ref{fig:sogs}. From the comparison of Figure~\ref{fig:sol1sog} and Figure~\ref{fig:sol2sog}, we can notice the discrepancy despite the two functions are solving the same problem.

In fact, going back to the creation of SOG, it was created to preserve semantic structures and eliminates semantics-independent elements. And that is why it was very effective in bypassing the mutation and optimizations the compiler does to the same instruction line. However, as stated in previous sections, this "semantic" preserving technique only works on the Low-level and Mid-level differences, where the source code is the same and the differences are created by doing mutations and alterations to the order and form of the instructions. In our Semantic Dataset, all the differences are created from the source code level, where each source code is written by a different person, but achieving the same goal. This is the high-level differences that HermesSim was not able to solve with its SOGs. And such high-level differences could potentially be exploited in adversarial settings to reduce the effectiveness of BFSD models and tools.

However, how does Trex \cite{Trex} still manage to achieve a relatively high AUC on the Semantic Dataset? We believe the cause stems from the dynamic analysis included in Trex’s binary analysis phase. Specifically, Trex generates short “micro‐traces” by instrumenting each function and collecting its runtime behavior over a small number of representative inputs. These micro‐traces capture semantic features, such as sequences of basic blocks executed, memory access patterns, or key API calls, that remain consistent across different source implementations of the same functionality. In contrast to purely static or token‐based embeddings, micro‐traces effectively abstract away syntactic differences introduced by varied source code or compilation settings, enabling Trex to match semantically equivalent functions even when their disassembly looks very different. Consequently, on the Semantic Dataset, Trex’s ability to leverage dynamic micro‐traces results in a noticeably higher AUC than models relying solely on static features, in this case HermesSim.

\vspace{-0.5\baselineskip}

\input{tables/d4analysis}

%% file: tables/d1results.tex
\begin{table*}[!t]
\centering
\caption{Inter-Model Comparison of Standard Dataset.}
\label{tab:d1comparison}
\scalebox{0.6}{
\begin{tblr}{
  columns = {c},
  column{1} = {l},
  cell{1}{1} = {r=2}{},
  cell{1}{2} = {c=8}{c},
  cell{1}{11} = {c=4}{},
  cell{1}{16} = {c=4}{},
  cell{1}{21} = {c=3}{},
  hline{1,3,12} = {-}{},
  hline{2} = {2-9,11-14,16-19,21-23}{},
}
Model Name                                 & AUC           &               &               &               &               &               &               &               &  & MRR10         &               &               &               &  & Recall@1      &               &               &               &  & Testing Time (s) &       &         &  \\
                                           & XA            & XB            & XA+XB         & XC            & XC+XB         & XV            & XO            & XM            &  & XA+XB         & XC            & XC+XB         & XM            &  & XA+XB         & XC            & XC+XB         & XM            &  & Feat             & Inf   & Tot 100 &  \\
FCatalog \cite{FCatalog}                   & 0.44          & 0.79          & 0.44          & 0.81          & 0.62          & 0.99          & 0.94          & 0.66          &  & 0.06          & 0.36          & 0.16          & 0.36          &  & 0.02          & 0.24          & 0.07          & 0.30          &  & 11840            & 0     & 3.83    &  \\
FunctionSimSearch \cite{functionsimsearch} & 0.72          & 0.83          & 0.72          & 0.66          & 0.65          & 0.81          & 0.79          & 0.72          &  & 0.26          & 0.20          & 0.15          & 0.32          &  & 0.19          & 0.15          & 0.10          & 0.26          &  & 8204             & 6526  & 4.76    &  \\
Asm2Vec \cite{Asm2Vec}                     & 0.44          & 0.66          & 0.44          & 0.71          & 0.62          & 0.81          & 0.77          & 0.70          &  & 0.01          & 0.18          & 0.02          & 0.15          &  & 0.00          & 0.10          & 0.00          & 0.11          &  & 2191             & 1382  & 1.15    &  \\
Massarelli et al. \cite{Massarelli}        & 0.87          & 0.93          & 0.87          & 0.80          & 0.78          & 0.95          & 0.90          & 0.88          &  & 0.29          & 0.25          & 0.16          & 0.41          &  & 0.16          & 0.17          & 0.08          & 0.31          &  & 299              & 5404  & 1.84    &  \\
Gemini \cite{gemini}                       & 0.96          & 0.98          & 0.96          & 0.80          & 0.79          & 0.98          & 0.91          & 0.90          &  & 0.69          & 0.31          & 0.27          & 0.56          &  & 0.58          & 0.24          & 0.19          & 0.47          &  & 2657             & 1004  & 1.18    &  \\
SAFE \cite{SAFE}                           & 0.79          & 0.90          & 0.76          & 0.83          & 0.78          & 0.93          & 0.92          & 0.86          &  & 0.16          & 0.22          & 0.11          & 0.35          &  & 0.09          & 0.11          & 0.04          & 0.25          &  & 160              & 236   & 0.13    &  \\
Zeek \cite{Zeek}                           & 0.92          & 0.96          & 0.91          & 0.82          & 0.81          & 0.96          & 0.89          & 0.91          &  & 0.44          & 0.24          & 0.22          & 0.41          &  & 0.30          & 0.16          & 0.13          & 0.27          &  & 93816            & 62    & 30.33   &  \\
Trex \cite{Trex}                           & 0.89          & 0.97          & 0.89          & 0.81          & 0.76          & 0.99          & 0.95          & 0.89          &  & 0.47          & 0.36          & 0.24          & 0.57          &  & 0.35          & 0.30          & 0.17          & 0.50          &  & 766              & 35146 & 11.60   &  \\
HermesSim \cite{HermesSim}                 & \textbf{1.00} & \textbf{1.00} & \textbf{1.00} & \textbf{0.97} & \textbf{0.96} & \textbf{1.00} & \textbf{1.00} & \textbf{0.99} &  & \textbf{0.99} & \textbf{0.77} & \textbf{0.68} & \textbf{0.93} &  & \textbf{0.99} & \textbf{0.72} & \textbf{0.61} & \textbf{0.91} &  & 3881             & 4765  & 2.79    &  
\end{tblr}
}
\captionsetup{justification=centering, margin=0.5cm}
\caption*{\small Feat: feature extraction; Inf: inference; Tot 100: the combined time used per 100 functions on average.}
\end{table*}

%% file: tables/d2results.tex
\begin{table}[!b]
\centering
\caption{Inter-Model Comparison of Firmware Dataset.}
\label{tab:d2comparison}
\scalebox{0.7}{
\begin{tblr}{
  cell{1}{1} = {r=2}{},
  cell{1}{2} = {c=3}{c},
  cell{1}{6} = {c=3}{c},
  hline{1,3,12} = {-}{},
  hline{2} = {2-4,6-8}{},
}
Model Name                                 & AUC           &               &               &  & Testing Time (s) &      &         &  \\
                                           & XV            & XO            & XV+XO         &  & Feat             & Inf  & Tot 100 &  \\
FCatalog \cite{FCatalog}                   & 0.99          & 0.90          & 0.91          &  & 257              & 0    & 3.34    &  \\
FunctionSimSearch \cite{functionsimsearch} & 0.74          & 0.61          & 0.63          &  & 106              & 28   & 0.37    &  \\
Asm2Vec \cite{Asm2Vec}                     & 0.62          & 0.57          & 0.57          &  & 11               & 11   & 0.29    &  \\
Massarelli et al. \cite{Massarelli}        & 0.91          & 0.80          & 0.81          &  & 3                & 330  & 4.33    &  \\
Gemini \cite{gemini}                       & 0.96          & 0.85          & 0.87          &  & 20               & 105  & 1.62    &  \\
SAFE \cite{SAFE}                           & 0.95          & 0.83          & 0.85          &  & 2                & 33   & 0.45    &  \\
Zeek \cite{Zeek}                           & 0.82          & 0.69          & 0.70          &  & 196              & 11   & 2.69    &  \\
Trex \cite{Trex}                           & 0.90          & 0.75          & 0.79          &  & 10               & 6284 & 81.78   &  \\
HermesSim \cite{HermesSim}                 & \textbf{1.00} & \textbf{0.99} & \textbf{0.99} &  & 765              & 2084 & 37.02   &  
\end{tblr}
}
\end{table}

%% file: tables/d3results.tex
\begin{table}[!b]
\centering
\caption{Inter-Model Comparison of Malware Dataset.}
\label{tab:d3comparison}
\scalebox{0.7}{
\begin{tblr}{
  column{even} = {c},
  column{3} = {c},
  column{7} = {r},
  cell{1}{1} = {r=2}{},
  cell{1}{2} = {c=3}{},
  cell{1}{6} = {c=3}{},
  cell{2}{7} = {c},
  cell{3}{6} = {r},
  cell{3}{8} = {r},
  cell{4}{6} = {r},
  cell{4}{8} = {r},
  cell{5}{6} = {r},
  cell{5}{8} = {r},
  cell{6}{6} = {r},
  cell{6}{8} = {r},
  cell{7}{6} = {r},
  cell{7}{8} = {r},
  cell{8}{6} = {r},
  cell{8}{8} = {r},
  cell{9}{6} = {r},
  cell{9}{8} = {r},
  cell{10}{6} = {r},
  cell{10}{8} = {r},
  cell{11}{6} = {r},
  cell{11}{8} = {r},
  hline{1,3,12} = {-}{},
  hline{2} = {2-4,6-8}{},
}
Model Name                                 & AUC           &               &               &  & Testing Time (s) &       &         &  \\
                                           & XC            & XO            & XV            &  & Feat             & Inf   & Tot 100 &  \\
FCatalog \cite{FCatalog}                   & 0.83          & 0.89          & \textbf{0.99} &  & 352              & 0     & 24.50   &  \\
FunctionSimSearch \cite{functionsimsearch} & 0.8           & 0.9           & 0.97          &  & 214              & 39    & 17.61   &  \\
Asm2Vec\cite{Asm2Vec}                      & 0.92          & 0.91          & \textbf{0.99} &  & 17               & 9     & 1.81    &  \\
Massarelli et al. \cite{Massarelli}        & 0.9           & 0.91          & 0.93          &  & 2                & 246   & 17.26   &  \\
Gemini \cite{gemini}                       & 0.89          & \textbf{0.95} & \textbf{0.99} &  & 31               & 37    & 4.73    &  \\
SAFE\cite{SAFE}                            & \textbf{0.95} & \textbf{0.95} & 0.97          &  & 2                & 35    & 2.57    &  \\
Zeek\cite{Zeek}                            & 0.94          & 0.94          & 0.98          &  & 71               & 9     & 5.57    &  \\
Trex\cite{Trex}                            & 0.92          & 0.92          & \textbf{0.99} &  & 5                & 6,060 & 422.06  &  \\
HermesSim\cite{HermesSim}                  & \textbf{0.95} & \textbf{0.95} & 0.98          &  & 1,176            & 3,511 & 326.17  &  
\end{tblr}
}
\end{table}

%% file: tables/d4results.tex
\begin{table}[!t]
\centering
\caption{Inter-Model Comparison of Obfuscation Dataset.}
\label{tab:d4comparison}
\scalebox{0.65}{
\begin{tblr}{
  column{even} = {c},
  column{3} = {c},
  column{7} = {r},
  cell{1}{1} = {r=2}{},
  cell{1}{2} = {c=3}{},
  cell{1}{6} = {c=3}{},
  cell{2}{7} = {c},
  cell{3}{6} = {r},
  cell{3}{8} = {r},
  cell{4}{6} = {r},
  cell{4}{8} = {r},
  cell{5}{6} = {r},
  cell{5}{8} = {r},
  cell{6}{6} = {r},
  cell{6}{8} = {r},
  cell{7}{6} = {r},
  cell{7}{8} = {r},
  cell{8}{6} = {r},
  cell{8}{8} = {r},
  cell{9}{6} = {r},
  cell{9}{8} = {r},
  cell{10}{6} = {r},
  cell{10}{8} = {r},
  cell{11}{6} = {r},
  cell{11}{8} = {r},
  hline{1,3,12} = {-}{},
  hline{2} = {2-4,6-8}{},
}
Model Name                                 & XOb           &               &               &  & Testing Time (s) &       &         &  \\
                                           & AUC           & MRR@10        & Recall@1      &  & Feat             & Inf   & Tot 100 &  \\
FCatalog \cite{FCatalog}                   & 0.78          & 0.38          & 0.34          &  & 1,600            & 0     & 6.33    &  \\
FunctionSimSearch \cite{functionsimsearch} & 0.64          & 0.17          & 0.11          &  & 1,032            & 805   & 7.26    &  \\
Asm2Vec \cite{Asm2Vec}                     & 0.67          & 0.19          & 0.16          &  & 266              & 297   & 2.23    &  \\
Massarelli et al. \cite{Massarelli}        & 0.92          & 0.43          & 0.26          &  & 70               & 1,630 & 6.72    &  \\
Gemini \cite{gemini}                       & 0.84          & 0.25          & 0.16          &  & 3,024            & 204   & 12.76   &  \\
SAFE \cite{SAFE}                           & 0.92          & 0.46          & 0.46          &  & 35               & 48    & 0.33    &  \\
Zeek \cite{Zeek}                           & 0.93          & 0.40          & 0.40          &  & 613              & 18    & 2.50    &  \\
Trex \cite{Trex}                           & 0.75          & 0.32          & 0.32          &  & 159              & 9,713 & 39.04   &  \\
HermesSim \cite{HermesSim}                 & \textbf{1.00} & \textbf{1.00} & \textbf{1.00} &  & 3,289            & 5,113 & 33.22   &  
\end{tblr}
}
\end{table}

%% file: tables/d5results.tex
\begin{table}[!b]
\centering
\caption{Inter-Model Comparison of Semantic Dataset.}
\label{tab:d5comparison}
\scalebox{0.65}{
\begin{tblr}{
  column{even} = {c},
  column{3} = {c},
  column{7} = {r},
  cell{1}{1} = {r=2}{},
  cell{1}{2} = {c=3}{},
  cell{1}{6} = {c=3}{},
  cell{2}{7} = {c},
  cell{3}{6} = {r},
  cell{3}{8} = {r},
  cell{4}{6} = {r},
  cell{4}{8} = {r},
  cell{5}{6} = {r},
  cell{5}{8} = {r},
  cell{6}{6} = {r},
  cell{6}{8} = {r},
  cell{7}{6} = {r},
  cell{7}{8} = {r},
  cell{8}{6} = {r},
  cell{8}{8} = {r},
  cell{9}{6} = {r},
  cell{9}{8} = {r},
  cell{10}{6} = {r},
  cell{10}{8} = {r},
  cell{11}{6} = {r},
  cell{11}{8} = {r},
  hline{1,3,12} = {-}{},
  hline{2} = {2-4,6-8}{},
}
Model Name                                 & XS            &               &               &  & Testing Time (s) &        &         &  \\
                                           & AUC           & Recall@1      & MRR@10        &  & Feat             & Inf    & Tot 100 &  \\
FCatalog \cite{FCatalog}                   & 0.70          & 0.27          & 0.28          &  & 627              & 0      & 50.12   &  \\
FunctionSimSearch \cite{functionsimsearch} & 0.68          & 0.18          & 0.16          &  & 657              & 9      & 53.24   &  \\
Asm2Vec \cite{Asm2Vec}                     & 0.53          & 0.17          & 0.11          &  & 5                & 5      & 0.80    &  \\
Massarelli et al. \cite{Massarelli}        & 0.73          & 0.14          & 0.11          &  & 2                & 728    & 58.35   &  \\
Gemini \cite{gemini}                       & 0.83          & 0.33          & 0.30          &  & 3                & 138    & 11.27   &  \\
SAFE \cite{SAFE}                           & \textbf{0.86} & 0.30          & 0.28          &  & 1                & 86     & 6.95    &  \\
Zeek \cite{Zeek}                           & 0.76          & 0.17          & 0.14          &  & 26               & 21     & 3.77    &  \\
Trex \cite{Trex}                           & 0.82          & \textbf{0.34} & \textbf{0.33} &  & 2                & 23,503 & 1,878.90 &  \\
HermesSim \cite{HermesSim}                 & 0.70          & 0.21          & 0.20          &  & 1,345            & 12,235 & 1,085.53 &  
\end{tblr}
}
\end{table}

%% file: tables/d4analysis.tex
\begin{table}
\centering
\caption{AUC Comparison Within Obfuscation Dataset.}
\label{tab:d3comparison}
\scalebox{0.7}{
\begin{tblr}{
  column{3} = {c},
  column{4} = {c},
  cell{1}{1} = {c=2}{},
  cell{2}{1} = {c=2}{},
  cell{3}{1} = {c=2}{},
  cell{8}{1} = {c=2}{},
  cell{9}{1} = {c=2}{},
  cell{10}{1} = {c=2}{},
  cell{11}{1} = {c=2}{},
  cell{12}{1} = {c=2}{},
  cell{13}{1} = {c=2}{},
  cell{14}{1} = {c=2}{},
  hline{1-3,8,15} = {-}{},
}
Model Name                                 &                        & Tigress Obfuscations & LLVM Obfuscations \\
FCatalog \cite{FCatalog}                   &                        & 1.00                 & 0.81              \\
FunctionSimSearch \cite{functionsimsearch} &                        &                      &                   \\
                                           & G                      & 0.98                 & 0.56              \\
                                           & G + M                  & 1.00                 & 0.64              \\
                                           & G + M + I              & 1.00                 & 0.65              \\
                                           & \textit{w} (G + M + I) & 0.67                 & 0.64              \\
Asm2Vec\cite{Asm2Vec}                      &                        & 1.00                 & 0.68              \\
Massarelli et al. \cite{Massarelli}        &                        & 1.00                 & 0.93              \\
Gemini \cite{gemini}                       &                        & 0.98                 & 0.84              \\
SAFE\cite{SAFE}                            &                        & 1.00                 & 0.93              \\
Zeek\cite{Zeek}                            &                        & 1.00                 & 0.94              \\
Trex\cite{Trex}                            &                        & 1.00                 & 0.75              \\
HermesSim\cite{HermesSim}                  &                        & 1.00                 & 1.00              
\end{tblr}
}
\end{table}

%% file: paper/05_discussion.tex
\paragraph{Robustness Across Variation Levels.}
Although HermesSim~\cite{HermesSim} achieves the strongest overall performance and maintains AUC above $0.92$ across most datasets, its performance degrades under high-level semantic variation. More broadly, no single model consistently performs well across all variation levels, indicating that BFSD robustness varies substantially with the type of binary variation.

\paragraph{Model Trends.}
Graph-based methods appear to be the most promising for solving the BFSD problem. Models such as HermesSim and Gemini consistently outperform others in terms of accuracy metrics like AUC, Recall, and MRR. Their advantage primarily lies in the ability of graph representations, such as ACFG and SOG, to effectively model semantic relationships and structural patterns within binary functions. Meanwhile, NLP-based models show potential but generally lag behind graph-based methods, especially when dealing with obfuscation and semantic variability.

\paragraph{Dataset Effectiveness.}
Most datasets effectively highlight distinct evaluation criteria and provide comprehensive coverage for robustness evaluation. The Standard Dataset benchmarks conventional scenarios, while the Firmware Dataset exposes limitations in handling embedded binaries. The Malware Dataset shows limited differentiation due to challenges in dataset construction. The Obfuscation and Semantic Datasets further challenge models by introducing transformation diversity and semantic variations.

\paragraph{Future Benchmark Directions.}
Expanding dataset diversity and complexity remains important. Future benchmarks could incorporate more programming languages, advanced obfuscation techniques, diverse binary formats, and a broader range of real-world binaries to provide more comprehensive evaluations of BFSD robustness.

\paragraph{Directions for Future Benchmark Development.}
Future datasets should aim to address existing gaps, including:
\begin{itemize}
\item Increased semantic variability by including diverse sources and human-written functions with nuanced semantic differences.
\item Expanded obfuscation coverage, incorporating state-of-the-art obfuscation methods that are actively used in industry and cybersecurity practices.
\item Inclusion of diverse binary formats and languages to better evaluate cross-platform and cross-architecture capabilities.
\item Creation of larger datasets with realistic scales, ensuring that performance metrics like efficiency (feature extraction and inference time) become critical and practically relevant criteria for model assessment.
\end{itemize}

By addressing these improvements, future research can substantially enhance the robustness and reliability of BFSD models, enabling better practical deployment.

%% file: paper/06_conclusion.tex
In this paper, we introduced \sys, a benchmark specifically designed to provide realistic and diverse evaluation scenarios for Binary Function Similarity Detection (BFSD). Through comprehensive evaluations on five carefully constructed datasets, each targeting different critical aspects of binary function variability, we revealed significant insights into the strengths and weaknesses of current state-of-the-art models. Our results demonstrate the importance of graph-based methods for achieving robust and reliable performance, highlighting HermesSim \cite{HermesSim} and Gemini \cite{gemini} as particularly promising solutions. Additionally, our analysis underscores the need for more diverse and representative datasets that include real-world binaries, advanced obfuscation techniques, and nuanced semantic variations. We believe that addressing these dataset limitations and methodological challenges will substantially enhance the efficacy and generalizability of BFSD models, paving the way for their successful application in real-world software security tasks.

%% file: paper/07_ethics.tex
This research involved the construction and analysis of datasets composed of binary code and corresponding source code collected from publicly accessible open-source projects, vendor-distributed firmware images, public malware repositories, and programming contest submissions.

For open-source software and vendor-released source code, we systematically reviewed the associated licenses and included only artifacts whose terms permit redistribution for research purposes. All firmware binaries included in the dataset were obtained from vendor-provided example firmware packages or publicly downloadable firmware images. These artifacts were originally distributed by the respective manufacturers. We did not circumvent access controls, extract proprietary components from restricted devices, or modify firmware beyond what was required for analysis and compilation.

Malware samples were collected from publicly accessible repositories (e.g., publicly available GitHub projects) that host historical malware code for research and archival purposes. These samples were already publicly distributed at the time of collection. We did not create, enhance, or operationalize malicious capabilities, and no additional exploit functionality was introduced. The dataset is intended strictly for reproducible defensive research.

For programming contest submissions (e.g., AtCoder), we do not redistribute original source code or compiled binaries. Instead, we provide data collection scripts that retrieve submissions directly from the platform in accordance with its public access mechanisms, and we release only derived feature representations that are not reversible to the original code. No personal data is included in the dataset.

All experiments involving malware and firmware were conducted in secure, isolated computing environments with restricted network connectivity to prevent unintended propagation or interaction with external systems. This research complies with applicable institutional guidelines and legal standards and is conducted solely to advance defensive and scientific research in binary function similarity detection.

%% file: paper/08_acknowledgments.tex
The authors would like to acknowledge Shivam Parikh for his assistance with the data collection of the semantic dataset in the early stages of this project..

%% file: paper/09_availability.tex
All datasets in \sys as well as the collected and implemented models will be released in anonymized form for the review process.

\noindent\textbf{Code and scripts.}
The code and scripts used in this paper can be found on Github:
\url{https://github.com/fan1192/EXHIB}

\noindent\textbf{Binaries and extracted features.}
We host the binaries and extracted features on Zenodo:
\url{https://doi.org/10.5281/zenodo.18488936}

\noindent\textbf{Semantic dataset.}
For the programming contest submissions (semantic dataset), we do not redistribute the original source code or compiled binaries, as these remain subject to the platform’s terms of service. Instead, we release only the extracted feature representations used in our experiments, along with data collection scripts that allow researchers to retrieve the submissions directly from the official platform through its publicly accessible interfaces.

\noindent\textbf{Integrity.}
The Zenodo record includes SHA-256 checksums for all released archives.

%% file: myReferences.bib
@inproceedings{CLAP,
author = {Wang, Hao and Gao, Zeyu and Zhang, Chao and Sha, Zihan and Sun, Mingyang and Zhou, Yuchen and Zhu, Wenyu and Sun, Wenju and Qiu, Han and Xiao, Xi},
title = {CLAP: Learning Transferable Binary Code Representations with Natural Language Supervision},
year = {2024},
isbn = {9798400706127},
publisher = {Association for Computing Machinery},
address = {New York, NY, USA},
url = {https://doi.org/10.1145/3650212.3652145},
doi = {10.1145/3650212.3652145},
booktitle = {Proceedings of the 33rd ACM SIGSOFT International Symposium on Software Testing and Analysis},
pages = {503–515},
numpages = {13},
location = {Vienna, Austria},
series = {ISSTA 2024}
}

@inproceedings{Co2FuLL,
author = {Dong, Chaopeng and Guo, Jingdong and Yang, Shouguo and Li, Yi and Fang, Dongliang and Xiao, Yang and Chen, Yongle and Sun, Limin},
title = {Advancing Binary Code Similarity Detection via Context-Content Fusion and LLM Verification},
year = {2025},
publisher = {IEEE Press},
url = {https://doi.org/10.1109/ASE63991.2025.00033},
doi = {10.1109/ASE63991.2025.00033},
booktitle = {2025 40th IEEE/ACM International Conference on Automated Software Engineering (ASE)},
pages = {304–316},
numpages = {13},
location = {Seoul, Korea, Republic of}
}

@InProceedings{Roussev2010,
author="Roussev, Vassil",
editor="Chow, Kam-Pui
and Shenoi, Sujeet",
title="Data Fingerprinting with Similarity Digests",
booktitle="Advances in Digital Forensics VI",
year="2010",
publisher="Springer Berlin Heidelberg",
address="Berlin, Heidelberg",
pages="207--226",
isbn="978-3-642-15506-2"
}

@article{Kornblum2006,
title = {Identifying almost identical files using context triggered piecewise hashing},
journal = {Digital Investigation},
volume = {3},
pages = {91-97},
year = {2006},
note = {The Proceedings of the 6th Annual Digital Forensic Research Workshop (DFRWS '06)},
issn = {1742-2876},
doi = {https://doi.org/10.1016/j.diin.2006.06.015},
url = {https://www.sciencedirect.com/science/article/pii/S1742287606000764},
author = {Jesse Kornblum}
}

@INPROCEEDINGS{9110432,
  author={Bak, Márton and Papp, Dorottya and Tamás, Csongor and Buttyán, Levente},
  booktitle={NOMS 2020 - 2020 IEEE/IFIP Network Operations and Management Symposium}, 
  title={Clustering IoT Malware based on Binary Similarity}, 
  year={2020},
  volume={},
  number={},
  pages={1-6},
  doi={10.1109/NOMS47738.2020.9110432}}

@inproceedings{Genealogy,
author = {Cozzi, Emanuele and Vervier, Pierre-Antoine and Dell'Amico, Matteo and Shen, Yun and Bilge, Leyla and Balzarotti, Davide},
title = {The Tangled Genealogy of IoT Malware},
year = {2020},
isbn = {9781450388580},
publisher = {Association for Computing Machinery},
address = {New York, NY, USA},
url = {https://doi.org/10.1145/3427228.3427256},
doi = {10.1145/3427228.3427256},
booktitle = {Proceedings of the 36th Annual Computer Security Applications Conference},
pages = {1–16},
numpages = {16},
location = {Austin, USA},
series = {ACSAC '20}
}

@INPROCEEDINGS{Vuddy,
  author={Kim, Seulbae and Woo, Seunghoon and Lee, Heejo and Oh, Hakjoo},
  booktitle={2017 IEEE Symposium on Security and Privacy (SP)}, 
  title={VUDDY: A Scalable Approach for Vulnerable Code Clone Discovery}, 
  year={2017},
  volume={},
  number={},
  pages={595-614},
  doi={10.1109/SP.2017.62}}

@INPROCEEDINGS{ReDeBug,
  author={Jang, Jiyong and Agrawal, Abeer and Brumley, David},
  booktitle={2012 IEEE Symposium on Security and Privacy}, 
  title={ReDeBug: Finding Unpatched Code Clones in Entire OS Distributions}, 
  year={2012},
  volume={},
  number={},
  pages={48-62},
  doi={10.1109/SP.2012.13}}

@inproceedings{tigress,
  title={Tigress: A source-to-source-ish obfuscation tool},
  author={Collberg, Christian},
  booktitle={Proc. 8th Workshop on Software Security, Protection, and Reverse Engineering},
  year={2018}
}

@misc{radare2,
  title        = {{Radare2}},
  author       = {{Radare2 Project}},
  howpublished = {\url{https://rada.re/n/}},
  year         = {2025}
}

@misc{bindiff,
  title        = {BinDiff},
  author       = {{Zynamics (acquired by Google)}},
  howpublished = {\url{https://www.zynamics.com/bindiff.html}},
  year         = {2025}
}

@inproceedings{binmatch,
  title     = {BinMatch: A Semantics-Based Hybrid Approach on Binary Code Clone Analysis},
  author    = {Yikun Hu and Yuanyuan Zhang and Juanru Li and Hui Wang and Bodong Li and Dawu Gu},
  booktitle = {Proceedings of the 25th IEEE International Conference on Software Analysis, Evolution and Reengineering (SANER)},
  year      = {2018},
  pages     = {1--12},
  doi       = {10.1109/SANER.2018.8330203},
  url       = {https://arxiv.org/abs/1808.06216}
}

@article{malpedia,
  title     = {Malpedia: A Collaborative Effort to Inventorize the Malware Landscape},
  author    = {Plohmann, Daniel and Clau{\ss}, Martin and Enders, Steffen and Padilla, Elmar},
  journal   = {The Journal on Cybercrime and Digital Investigations},
  volume    = {3},
  number    = {1},
  pages     = {1--19},
  year      = {2017},
  publisher = {CECyF},
  doi       = {10.18464/cybin.v3i1.17},
  url       = {https://cyberjournal.cecyf.fr/index.php/cybin/article/view/17}
}

@misc{binpro,
      title={BinPro: A Tool for Binary Source Code Provenance}, 
      author={Dhaval Miyani and Zhen Huang and David Lie},
      year={2017},
      eprint={1711.00830},
      archivePrefix={arXiv},
      primaryClass={cs.CR},
      url={https://arxiv.org/abs/1711.00830}, 
}

@inproceedings{deepbindiff,
  title     = {DeepBinDiff: Learning Program-Wide Code Representations for Binary Diffing},
  author    = {Yue Duan and Xuezixiang Li and Jinghan Wang and Heng Yin},
  booktitle = {Proceedings of the Network and Distributed System Security Symposium (NDSS)},
  year      = {2020},
  doi       = {10.14722/ndss.2020.24311},
  url       = {https://www.ndss-symposium.org/wp-content/uploads/2020/02/24311.pdf}
}

@inproceedings{Kam1n0,
author = {Ding, Steven H.H. and Fung, Benjamin C.M. and Charland, Philippe},
title = {Kam1n0: MapReduce-based Assembly Clone Search for Reverse Engineering},
year = {2016},
isbn = {9781450342322},
publisher = {Association for Computing Machinery},
address = {New York, NY, USA},
url = {https://doi.org/10.1145/2939672.2939719},
doi = {10.1145/2939672.2939719},
booktitle = {Proceedings of the 22nd ACM SIGKDD International Conference on Knowledge Discovery and Data Mining},
pages = {461–470},
numpages = {10},
location = {San Francisco, California, USA},
series = {KDD '16}
}

@misc{ghidra,
  author       = {{National Security Agency}},
  title        = {Ghidra Software Reverse Engineering Framework},
  year         = {2019},
  howpublished = {\url{https://ghidra-sre.org/}},
}

@misc{networkx,
  author       = {Aric A. Hagberg and Daniel A. Schult and Pieter J. Swart},
  title        = {NetworkX: Software for Complex Networks},
  howpublished = {\url{https://networkx.github.io/}},
  year         = {2008},
}

@misc{capstone,
  author       = {Nguyen Anh Quynh and others},
  title        = {Capstone: A Lightweight Multi-Platform, Multi-Architecture Disassembly Framework},
  howpublished = {\url{http://www.capstone-engine.org}},
  year         = {2013}
}

@misc{atcoder,
  author = {{AtCoder}},
  title = {AtCoder},
  howpublished = {\url{https://atcoder.jp/}},
}

@misc{hikari,
  author = {Naville Zhang},
  title = {Hikari – an improvement over Obfuscator-LLVM},
  year = {2017},
  howpublished = {\url{https://github.com/HikariObfuscator/Hikari}},

}

@INPROCEEDINGS{Obfuscator-LLVM,
  author={Pascal Junod and Julien Rinaldini and Johan Wehrli and Julie Michielin},
  booktitle={Proceedings of the {IEEE/ACM} 1st International Workshop on Software Protection, {SPRO'15}, Firenze, Italy, May 19th, 2015},
  editor = {Brecht Wyseur},
  publisher = {IEEE},
  title={Obfuscator-{LLVM} -- Software Protection for the Masses},
  year={2015},
  pages={3--9},
  doi={10.1109/SPRO.2015.10},
}

@misc{iot-malware-2017,
  author = {Fei Ding},
  title = {IoT Malware},
  year = {2017},
  howpublished = {\url{https://github.com/ifding/iot-malware}}
}

@inproceedings{ding2020deeppower,
  title={DeepPower: Non-intrusive and Deep Learning-based Detection of IoT Malware Using Power Side Channels},
  author={Ding, Fei and Li, Hongda and Luo, Feng and Hu, Hongxin and Cheng, Long and Xiao, Hai and Ge, Rong},
  booktitle={Proceedings of the 15th ACM Asia Conference on Computer and Communications Security},
  pages={33--46},
  year={2020}
}

@misc{vxunderground_malware,
  author       = {vxunderground},
  title        = {MalwareSourceCode},
  year         = {2023},
  howpublished = {\url{https://github.com/vxunderground/MalwareSourceCode}},
}

@software{nrf5_sdk,
  author    = {{Nordic Semiconductor}},
  title     = {nRF5 SDK},
  version   = {17.1.0},
  year      = {2023},
  url       = {https://developer.nordicsemi.com/nRF5_SDK/},
  note      = {Software Development Kit for Nordic Semiconductor nRF devices},
}

@software{SimplicityStudio,
  author    = {{Silicon Labs}},
  title     = {Simplicity Studio},
  version   = {5.5},
  year      = {2023},
  url       = {https://www.silabs.com/developers/simplicity-studio},
  note      = {Development Platform for Silicon Labs Devices},
}

@software{KeilMDK,
  author    = {{ARM Ltd.}},
  title     = {Keil MDK},
  version   = {5.38},
  year      = {2023},
  url       = {https://developer.arm.com/Tools%20and%20Software/Keil%20MDK},
  note      = {ARM Development Tools for Microcontrollers},
}

@InProceedings{lielal,
  title = 	 {Graph Matching Networks for Learning the Similarity of Graph Structured Objects},
  author =       {Li, Yujia and Gu, Chenjie and Dullien, Thomas and Vinyals, Oriol and Kohli, Pushmeet},
  booktitle = 	 {Proceedings of the 36th International Conference on Machine Learning},
  pages = 	 {3835--3845},
  year = 	 {2019},
  editor = 	 {Chaudhuri, Kamalika and Salakhutdinov, Ruslan},
  volume = 	 {97},
  series = 	 {Proceedings of Machine Learning Research},
  month = 	 {09--15 Jun},
  publisher =    {PMLR},
  url = 	 {https://proceedings.mlr.press/v97/li19d.html}
}

@article{Massarelli,
  title={Investigating Graph Embedding Neural Networks with Unsupervised Features Extraction for Binary Analysis},
  author={Luca Massarelli and Giuseppe Antonio Di Luna and Fabio Petroni and Leonardo Querzoni and Roberto Baldoni},
  journal={Proceedings 2019 Workshop on Binary Analysis Research},
  year={2019},
  url={https://api.semanticscholar.org/CorpusID:160018518}
}

@inproceedings{jTrans,
author = {Wang, Hao and Qu, Wenjie and Katz, Gilad and Zhu, Wenyu and Gao, Zeyu and Qiu, Han and Zhuge, Jianwei and Zhang, Chao},
title = {jTrans: jump-aware transformer for binary code similarity detection},
year = {2022},
isbn = {9781450393799},
publisher = {Association for Computing Machinery},
address = {New York, NY, USA},
url = {https://doi.org/10.1145/3533767.3534367},
doi = {10.1145/3533767.3534367},
booktitle = {Proceedings of the 31st ACM SIGSOFT International Symposium on Software Testing and Analysis},
pages = {1–13},
numpages = {13},
location = {Virtual, South Korea},
series = {ISSTA 2022}
}

@inproceedings{BERT,
  title={BERT: Pre-training of Deep Bidirectional Transformers for Language Understanding},
  author={Jacob Devlin and Ming-Wei Chang and Kenton Lee and Kristina Toutanova},
  booktitle={North American Chapter of the Association for Computational Linguistics},
  year={2019},
  url={https://api.semanticscholar.org/CorpusID:52967399}
}

@inproceedings{transformer,
author = {Vaswani, Ashish and Shazeer, Noam and Parmar, Niki and Uszkoreit, Jakob and Jones, Llion and Gomez, Aidan N. and Kaiser, \L{}ukasz and Polosukhin, Illia},
title = {Attention is all you need},
year = {2017},
isbn = {9781510860964},
publisher = {Curran Associates Inc.},
address = {Red Hook, NY, USA},
booktitle = {Proceedings of the 31st International Conference on Neural Information Processing Systems},
pages = {6000–6010},
numpages = {11},
location = {Long Beach, California, USA},
series = {NIPS'17}
}

@inproceedings{seq2seq,
author = {Sutskever, Ilya and Vinyals, Oriol and Le, Quoc V.},
title = {Sequence to sequence learning with neural networks},
year = {2014},
publisher = {MIT Press},
address = {Cambridge, MA, USA},
booktitle = {Proceedings of the 27th International Conference on Neural Information Processing Systems - Volume 2},
pages = {3104–3112},
numpages = {9},
location = {Montreal, Canada},
series = {NIPS'14}
}

@INPROCEEDINGS{Asm2Vec,
  author={Ding, Steven H. H. and Fung, Benjamin C. M. and Charland, Philippe},
  booktitle={2019 IEEE Symposium on Security and Privacy (SP)}, 
  title={Asm2Vec: Boosting Static Representation Robustness for Binary Clone Search against Code Obfuscation and Compiler Optimization}, 
  year={2019},
  volume={},
  number={},
  pages={472-489},
  doi={10.1109/SP.2019.00003}}

@inproceedings{Mikolov2013EfficientEO,
  title={Efficient Estimation of Word Representations in Vector Space},
  author={Tomas Mikolov and Kai Chen and Gregory S. Corrado and Jeffrey Dean},
  booktitle={International Conference on Learning Representations},
  year={2013},
  url={https://api.semanticscholar.org/CorpusID:5959482}
}

@inproceedings{NIPS2013_9aa42b31,
 author = {Mikolov, Tomas and Sutskever, Ilya and Chen, Kai and Corrado, Greg S and Dean, Jeff},
 booktitle = {Advances in Neural Information Processing Systems},
 editor = {C.J. Burges and L. Bottou and M. Welling and Z. Ghahramani and K.Q. Weinberger},
 pages = {},
 publisher = {Curran Associates, Inc.},
 title = {Distributed Representations of Words and Phrases and their Compositionality},
 url = {https://proceedings.neurips.cc/paper_files/paper/2013/file/9aa42b31882ec039965f3c4923ce901b-Paper.pdf},
 volume = {26},
 year = {2013}
}

@inproceedings {HermesSim,
author = {Haojie He and Xingwei Lin and Ziang Weng and Ruijie Zhao and Shuitao Gan and Libo Chen and Yuede Ji and Jiashui Wang and Zhi Xue},
title = {Code is not Natural Language: Unlock the Power of {Semantics-Oriented} Graph Representation for Binary Code Similarity Detection},
booktitle = {33rd USENIX Security Symposium (USENIX Security 24)},
year = {2024},
isbn = {978-1-939133-44-1},
address = {Philadelphia, PA},
pages = {1759--1776},
url = {https://www.usenix.org/conference/usenixsecurity24/presentation/he-haojie},
publisher = {USENIX Association},
month = aug
}

@misc{functionsimsearch,
  author       = {Thomas Dullien},
  title        = {Searching statically-linked vulnerable library functions in executable code},
  year         = {2018},
  howpublished = {\url{https://projectzero.google/2018/12/searching-statically-linked-vulnerable.html}},
}

@inproceedings{10.1145/3176258.3176306,
author = {Pagani, Fabio and Dell'Amico, Matteo and Balzarotti, Davide},
title = {Beyond Precision and Recall: Understanding Uses (and Misuses) of Similarity Hashes in Binary Analysis},
year = {2018},
isbn = {9781450356329},
publisher = {Association for Computing Machinery},
address = {New York, NY, USA},
url = {https://doi.org/10.1145/3176258.3176306},
doi = {10.1145/3176258.3176306},
booktitle = {Proceedings of the Eighth ACM Conference on Data and Application Security and Privacy},
pages = {354–365},
numpages = {12},
location = {Tempe, AZ, USA},
series = {CODASPY '18}
}

@inproceedings{10.1145/3238147.3238199,
author = {Liu, Bingchang and Huo, Wei and Zhang, Chao and Li, Wenchao and Li, Feng and Piao, Aihua and Zou, Wei},
title = {$\alpha$Diff: cross-version binary code similarity detection with DNN},
year = {2018},
isbn = {9781450359375},
publisher = {Association for Computing Machinery},
address = {New York, NY, USA},
url = {https://doi.org/10.1145/3238147.3238199},
doi = {10.1145/3238147.3238199},
booktitle = {Proceedings of the 33rd ACM/IEEE International Conference on Automated Software Engineering},
pages = {667–678},
numpages = {12},
location = {Montpellier, France},
series = {ASE '18}
}

@inproceedings{Zeek,
author = {Shalev, Noam and Partush, Nimrod},
title = {Binary Similarity Detection Using Machine Learning},
year = {2018},
isbn = {9781450359931},
publisher = {Association for Computing Machinery},
address = {New York, NY, USA},
url = {https://doi.org/10.1145/3264820.3264821},
doi = {10.1145/3264820.3264821},
booktitle = {Proceedings of the 13th Workshop on Programming Languages and Analysis for Security},
pages = {42–47},
numpages = {6},
location = {Toronto, Canada},
series = {PLAS '18}
}

@INPROCEEDINGS{10376642,
  author={Lin, Yijie and Yang, Mouxing and Yu, Jun and Hu, Peng and Zhang, Changqing and Peng, Xi},
  booktitle={2023 IEEE/CVF International Conference on Computer Vision (ICCV)}, 
  title={Graph Matching with Bi-level Noisy Correspondence}, 
  year={2023},
  volume={},
  number={},
  pages={23305-23314},
  doi={10.1109/ICCV51070.2023.02135}
}

@inproceedings{BinGo,
author = {Chandramohan, Mahinthan and Xue, Yinxing and Xu, Zhengzi and Liu, Yang and Cho, Chia Yuan and Tan, Hee Beng Kuan},
title = {BinGo: cross-architecture cross-OS binary search},
year = {2016},
isbn = {9781450342186},
publisher = {Association for Computing Machinery},
address = {New York, NY, USA},
url = {https://doi.org/10.1145/2950290.2950350},
doi = {10.1145/2950290.2950350},
booktitle = {Proceedings of the 2016 24th ACM SIGSOFT International Symposium on Foundations of Software Engineering},
pages = {678–689},
numpages = {12},
location = {Seattle, WA, USA},
series = {FSE 2016}
}

@inproceedings{10.1145/3634737.3644996,
author = {Kim, Taewook and Hong, Seokhyun and Cho, Yeongpil},
title = {AIMFuzz: Automated Function-Level In-Memory Fuzzing on Binaries},
year = {2024},
isbn = {9798400704826},
publisher = {Association for Computing Machinery},
address = {New York, NY, USA},
url = {https://doi.org/10.1145/3634737.3644996},
doi = {10.1145/3634737.3644996},
booktitle = {Proceedings of the 19th ACM Asia Conference on Computer and Communications Security},
pages = {1510–1522},
numpages = {13},
location = {Singapore, Singapore},
series = {ASIA CCS '24}
}

@inproceedings{BinSim,
author = {Ming, Jiang and Xu, Dongpeng and Jiang, Yufei and Wu, Dinghao},
title = {BinSim: trace-based semantic binary diffing via system call sliced segment equivalence checking},
year = {2017},
isbn = {9781931971409},
publisher = {USENIX Association},
address = {USA},
booktitle = {Proceedings of the 26th USENIX Conference on Security Symposium},
pages = {253–270},
numpages = {18},
location = {Vancouver, BC, Canada},
series = {SEC'17}
}

@inproceedings{10.5555/3155562.3155608,
author = {Karg\'{e}n, Ulf and Shahmehri, Nahid},
title = {Towards robust instruction-level trace alignment of binary code},
year = {2017},
isbn = {9781538626849},
publisher = {IEEE Press},
booktitle = {Proceedings of the 32nd IEEE/ACM International Conference on Automated Software Engineering},
pages = {342–352},
numpages = {11},
location = {Urbana-Champaign, IL, USA},
series = {ASE '17}
}

@inproceedings{Blanket_execution,
author = {Egele, Manuel and Woo, Maverick and Chapman, Peter and Brumley, David},
title = {Blanket execution: dynamic similarity testing for program binaries and components},
year = {2014},
isbn = {9781931971157},
publisher = {USENIX Association},
address = {USA},
booktitle = {Proceedings of the 23rd USENIX Conference on Security Symposium},
pages = {303–317},
numpages = {15},
location = {San Diego, CA},
series = {SEC'14}
}

@inproceedings{gemini,
author = {Xu, Xiaojun and Liu, Chang and Feng, Qian and Yin, Heng and Song, Le and Song, Dawn},
title = {Neural Network-based Graph Embedding for Cross-Platform Binary Code Similarity Detection},
year = {2017},
isbn = {9781450349468},
publisher = {Association for Computing Machinery},
address = {New York, NY, USA},
url = {https://doi.org/10.1145/3133956.3134018},
doi = {10.1145/3133956.3134018},
booktitle = {Proceedings of the 2017 ACM SIGSAC Conference on Computer and Communications Security},
pages = {363–376},
numpages = {14},
location = {Dallas, Texas, USA},
series = {CCS '17}
}

@inproceedings{10.1145/2976749.2978370,
author = {Feng, Qian and Zhou, Rundong and Xu, Chengcheng and Cheng, Yao and Testa, Brian and Yin, Heng},
title = {Scalable Graph-based Bug Search for Firmware Images},
year = {2016},
isbn = {9781450341394},
publisher = {Association for Computing Machinery},
address = {New York, NY, USA},
url = {https://doi.org/10.1145/2976749.2978370},
doi = {10.1145/2976749.2978370},
booktitle = {Proceedings of the 2016 ACM SIGSAC Conference on Computer and Communications Security},
pages = {480–491},
numpages = {12},
location = {Vienna, Austria},
series = {CCS '16}
}

@inproceedings{discovRE,
  title={discovRE: Efficient Cross-Architecture Identification of Bugs in Binary Code},
  author={Sebastian Eschweiler and Khaled Yakdan and Elmar Gerhards-Padilla},
  booktitle={Network and Distributed System Security Symposium},
  year={2016},
  url={https://api.semanticscholar.org/CorpusID:17208658}
}

@article{BinSequence,
  title={BinSequence: Fast, Accurate and Scalable Binary Code Reuse Detection},
  author={He Huang and Amr M. Youssef and Mourad Debbabi},
  journal={Proceedings of the 2017 ACM on Asia Conference on Computer and Communications Security},
  year={2017},
  url={https://api.semanticscholar.org/CorpusID:10493765}
}

@article{Yu_Cao_Tang_Nie_Huang_Wu_2020, title={Order Matters: Semantic-Aware Neural Networks for Binary Code Similarity Detection}, volume={34}, url={https://ojs.aaai.org/index.php/AAAI/article/view/5466}, DOI={10.1609/aaai.v34i01.5466}, abstractNote={&lt;p&gt;Binary code similarity detection, whose goal is to detect similar binary functions without having access to the source code, is an essential task in computer security. Traditional methods usually use graph matching algorithms, which are slow and inaccurate. Recently, neural network-based approaches have made great achievements. A binary function is first represented as an control-flow graph (CFG) with manually selected block features, and then graph neural network (GNN) is adopted to compute the graph embedding. While these methods are effective and efficient, they could not capture enough semantic information of the binary code. In this paper we propose semantic-aware neural networks to extract the semantic information of the binary code. Specially, we use BERT to pre-train the binary code on one token-level task, one block-level task, and two graph-level tasks. Moreover, we find that the order of the CFG’s nodes is important for graph similarity detection, so we adopt convolutional neural network (CNN) on adjacency matrices to extract the order information. We conduct experiments on two tasks with four datasets. The results demonstrate that our method outperforms the state-of-art models.&lt;/p&gt;}, number={01}, journal={Proceedings of the AAAI Conference on Artificial Intelligence}, author={Yu, Zeping and Cao, Rui and Tang, Qiyi and Nie, Sen and Huang, Junzhou and Wu, Shi}, year={2020}, month={Apr.}, pages={1145-1152} }

@inproceedings{VulSeeker,
author = {Gao, Jian and Yang, Xin and Fu, Ying and Jiang, Yu and Sun, Jiaguang},
title = {VulSeeker: a semantic learning based vulnerability seeker for cross-platform binary},
year = {2018},
isbn = {9781450359375},
publisher = {Association for Computing Machinery},
address = {New York, NY, USA},
url = {https://doi.org/10.1145/3238147.3240480},
doi = {10.1145/3238147.3240480},
booktitle = {Proceedings of the 33rd ACM/IEEE International Conference on Automated Software Engineering},
pages = {896–899},
numpages = {4},
location = {Montpellier, France},
series = {ASE '18}
}

@inproceedings{10.1145/3052973.3052995,
author = {Feng, Qian and Wang, Minghua and Zhang, Mu and Zhou, Rundong and Henderson, Andrew and Yin, Heng},
title = {Extracting Conditional Formulas for Cross-Platform Bug Search},
year = {2017},
isbn = {9781450349444},
publisher = {Association for Computing Machinery},
address = {New York, NY, USA},
url = {https://doi.org/10.1145/3052973.3052995},
doi = {10.1145/3052973.3052995},
booktitle = {Proceedings of the 2017 ACM on Asia Conference on Computer and Communications Security},
pages = {346–359},
numpages = {14},
location = {Abu Dhabi, United Arab Emirates},
series = {ASIA CCS '17}
}

@inproceedings{SAFE,
  title={SAFE: Self-Attentive Function Embeddings for Binary Similarity},
  author={Massarelli, Luca and Di Luna, Giuseppe Antonio and Petroni, Fabio and Querzoni, Leonardo and Baldoni, Roberto},
  booktitle={Proceedings of 16th Conference on Detection of Intrusions and Malware \& Vulnerability Assessment (DIMVA)},
  year={2019}
}

@misc{Trex,
      title={Trex: Learning Execution Semantics from Micro-Traces for Binary Similarity}, 
      author={Kexin Pei and Zhou Xuan and Junfeng Yang and Suman Jana and Baishakhi Ray},
      year={2021},
      eprint={2012.08680},
      archivePrefix={arXiv},
      primaryClass={cs.CR},
      url={https://arxiv.org/abs/2012.08680}, 
}

@misc{IDAPro,
  title = {IDA Pro: Interactive DisAssembler},
  author = {{Hex-Rays SA}},
  howpublished = {\url{https://hex-rays.com/ida-pro/}},
  year = {2024},
}

@misc{FCatalog,
  title = {The functions catalog},
  author = {Xorpd},
  howpublished = {\url{https://www.xorpd.net/pages/fcatalog.html}},
}

@article{10.1145/3446371,
author = {Haq, Irfan Ul and Caballero, Juan},
title = {A Survey of Binary Code Similarity},
year = {2021},
issue_date = {April 2022},
publisher = {Association for Computing Machinery},
address = {New York, NY, USA},
volume = {54},
number = {3},
issn = {0360-0300},
url = {https://doi.org/10.1145/3446371},
doi = {10.1145/3446371},
journal = {ACM Comput. Surv.},
month = {apr},
articleno = {51},
numpages = {38}
}

@inproceedings {usenix22,
author = {Andrea Marcelli and Mariano Graziano and Xabier Ugarte-Pedrero and Yanick Fratantonio and Mohamad Mansouri and Davide Balzarotti},
title = {How Machine Learning Is Solving the Binary Function Similarity Problem},
booktitle = {31st USENIX Security Symposium (USENIX Security 22)},
year = {2022},
isbn = {978-1-939133-31-1},
address = {Boston, MA},
pages = {2099--2116},
url = {https://www.usenix.org/conference/usenixsecurity22/presentation/marcelli},
publisher = {USENIX Association},
month = aug
}
